\documentclass[onecolumn]{IEEEtran}
\usepackage{mathtools}
\usepackage[verbose]{cite}
\usepackage[verbose]{wrapfig}
\usepackage[dvipsnames]{xcolor}
\usepackage{cellspace}
\usepackage{moresize}
\usepackage{epsfig,psfrag,subfigure}
\usepackage[font=scriptsize,farskip=5mm]{subfig}
\usepackage{pstricks-add}
\usepackage{amsmath,amssymb}
\usepackage{multirow,enumitem}

\newcommand{\mc}{\mathcal}
\newcommand{\mb}{\mathbf}

\newcommand{\mub}{\text{\boldmath${\mu}$}}
\newcommand{\rhob}{\text{\boldmath${\rho}$}}
\newcommand{\var}{\mathsf{var}}

\newtheorem{theorem}{Theorem}
\newtheorem{definition}{Definition}
\newtheorem{lemma}{Lemma}
\newtheorem{assumption}{Assumption}
\newtheorem{approximation}{Approximation}
\usepackage{nopageno}

\begin{document}
\thispagestyle{empty}

\IEEEoverridecommandlockouts
\title{A Characterization of Guesswork on\\Swiftly Tilting Curves}

\author{
	Ahmad Beirami,
	Robert Calderbank,
	Mark Christiansen,
	Ken Duffy,
	and Muriel M\'edard

\thanks{This work was presented in part in the 53rd Annual Allerton Conference on Communication, Control, and Computing (Allerton 2015)~\cite{Allerton15_guesswork}.}

\thanks{This work was supported in part by the Air Force Office of Scientific Reseah (AFOSR) under awards FA 9550-14-1-043 and FA 9550-13-1-0076;  the work of K. Duffy and M. M\'edard  was also supported in part by Netapp through a Faculty Fellowship. }

\thanks{A. Beirami {was with the Department of Electrical and Computer Engineering, Duke University, Durham, NC, USA, and also} with the Research Laboratory of Electronics, Massachusetts Institute of Technology, Cambridge, MA, USA. He is currently with EA Digital Platform -- Data \& AI, Electronic Arts, Redwood City, CA, USA (email: ahmad.beirami@gmail.com).}

\thanks{R. Calderbank is with the Department of Electrical and Computer Engineering, Duke University, {Durham, NC, }USA (email: robert.calderbank@duke.edu).}

\thanks{M. Christiansen was with the Hamilton Institute, Maynooth University, Ireland. He is currently with Grand Thornton (email: markchristiansen4224@gmail.com).}

\thanks{K. Duffy is with the Hamilton Institute, Maynooth University, Ireland (email: ken.duffy@mu.ie).}

\thanks{M. M\'edard is with the Research Laboratory of Electronics, Massachusetts Institute of Technology, Cambridge, MA, USA (email: medard@mit.edu).}

}

\maketitle

\begin{abstract}
Given a collection of strings, each with an associated probability of occurrence, the guesswork of each of them is their position in {a list ordered} from most likely to least likely, breaking ties arbitrarily.
Guesswork is central to several applications in information theory:
Average guesswork provides a lower bound on the expected computational cost of a sequential decoder to  decode successfully the transmitted message; the complementary cumulative distribution function of guesswork gives the error probability in list decoding; the logarithm of guesswork is the number of bits needed in optimal lossless one-to-one source coding; and guesswork is the number of trials required of an adversary to breach a password protected system in a brute-force attack.
In this paper, we consider memoryless string-sources 
that generate strings consisting of i.i.d. characters drawn from a finite alphabet,
and characterize their corresponding guesswork. Our main tool is the tilt operation on a memoryless string-source. We show that the tilt operation on a memoryless string-source parametrizes an exponential family of memoryless string-sources, which we refer to as the tilted family of the string-source.
We provide an operational meaning to the tilted families by proving that 
two memoryless string-sources result in the same guesswork on all strings of all lengths if and only if  their respective categorical distributions belong to the same tilted family.
Establishing some general properties of the tilt operation, we generalize the notions of weakly typical set and asymptotic equipartition property to tilted weakly typical sets of different orders. We use this new definition to characterize the large deviations for all atypical strings and characterize the volume of weakly typical sets of different orders. We subsequently build on this characterization to prove large deviation bounds on guesswork and provide an accurate approximation of its probability mass function. 
\end{abstract}


\section{Introduction}
\label{sec:intro}
\subsection{Motivation}

An early motivation for the study of guesswork is to provide lower bounds on the computational complexity of sequential decoding~\cite{sequential_decoding} where the decoder sequentially examines several paths until it finds the correct coded sequence.
Guesswork was first studied by Massey~\cite{massey94}, who showed that average guesswork is lower bounded in terms of the Shannon entropy of the source.  Ar{\i}kan~\cite{Arikan-guesswork} considered guesswork on memoryless sources and proved that the moments of guesswork are related to the R\'enyi entropy rates of the source in the  limit as the strings become long.
 This has been generalized to ergodic Markov chains~\cite{sullivan-markov}  and a wide range of stationary sources~\cite{sullivan-stationary}. 
 {
 Recently, tighter non-asymptotic bounds on the moments of guesswork were reported in~\cite{sason-guesswork}.}
Guesswork has also been studied subject to an allowable distortion~\cite{guesswork-distortion}, and subject to constrained Shannon entropy~\cite{ISIT15_guesswork}. 
Hanawal and Sundaresan~\cite{guessing-revisited} rederived the moments of guesswork assuming that the logarithm of guesswork satisfies a large deviations principle (LDP). 
Christiansen and Duffy~\cite{duffy-LDP} established that guesswork satisfies a LDP and characterized the rate function. They also provided an approximation to the distribution of guesswork using the rate function associated with the LDP.
{That LDP for guesswork was subsequently leveraged to provide a new proof of the classical Channel Coding Theorem and to 
analyze the performance of a novel, universal maximum likelihood decoding algorithm~\cite{Duffy-CCT}.}

In list decoding, rather than declaring one codeword message, the decoder returns a list of size $L$ of possible codewords~\cite{Elias-list-1, Elias-list-2, Sudan-list}. The error event in list decoding is the event where the message codeword is not returned in the list of the $L$ returned codewords. For  strings of fixed length, as the size of the list grows larger, the probability of error decreases.
The list decoder can be used as an inner code for an outer code that eventually decides the intended message from the returned list of messages by the inner list decoder. The optimal list decoder returns the $L$ most likely messages given the received codeword. Therefore, the probability of the error event in the optimal list decoder is characterized by the complementary cumulative distribution function of guesswork conditioned on the received codeword (see~\cite{Gallager-book,Merhav-list}).

Guesswork can also be used to quantify computational security against brute-force attack~\cite{multi-user}.
Suppose that a secret string is drawn from a given process on a finite alphabet, and is used to secure a password-protected system. Further suppose that the system only grants access if the correct secret string is provided exactly, and does not reveal any information otherwise. If a brute-force attacker adversary decides to guess the secret string by query, the best strategy at an adversary's disposal  is to query the possible secret strings from the most likely to the least likely (in terms of minimizing the average number of queries or maximizing the probability of success using a finite number of queries). In this case, guesswork gives the number of queries that it would take an attacker to find the secret string using the best strategy.
Christiansen {\em et al.}~\cite{ISIT13_Mark} studied guesswork over the weakly typical set and proved that the exponent is strictly smaller than that of a uniform set with the same support size; they showed that the average guesswork of a password over an erasure channel does not relate to the average noise in general~\cite{ASILOMAR13}; Christiansen {\em et al.}  also considered the setting where an attacker wants to guess one or more out of many secret strings drawn independently from not necessarily identical string-sources~\cite{multi-user}. Finally, the idea of guesswork has been extended to the setup where the probability distribution is unknown~\cite{sundaresan-universal, IT_Kosut_one2one, ISIT15_guesswork}.

In the context of source coding, it is known that the length of the optimal one-to-one source code  (that need not satisfy Kraft's inequality) is within one bit of the logarithm of guesswork. The average length of a one-to-one code is related to the normalized zeroth moment of guesswork or alternatively the expected logarithm of guesswork~(see \cite{IT_Kosut_one2one}). The source coding problem without prefix constraint dates back to Wyner~\cite{Wyner72}, who showed that the average codeword length of one-to-one codes is upper bounded by the entropy. Alon and Orlistky derived a lower bound on the average codeword length in terms of the Shannon entropy~\cite{Alon_Orlitsky_one2one}, which was recently revisited for other moments of  guesswork~\cite{courtade-verdu}. Szpankowski~\cite{Szpankowski_one2one} derived the asymptotic average codeword length of one-to-one codes on binary memoryless sources, which was subsequently generalized to finite-alphabet independent and identically distributed (i.i.d.) processes~\cite{Szpankowski2011,Kontoyiannis_one2one}, and later studied under a universal setup~\cite{IT_Kosut_one2one,ITW14}.

\subsection{Problem setup, definitions, and notation}
Let  $\mc{X} := (a_1, \ldots, a_{|\mc{X}|})$ be an ordered set that denotes a finite alphabet of size $|\mc{X}|$. We will subsequently use the ordering on $\mc{X}$ to break ties for equally likely strings in guesswork (and reverse guesswork). Denote an $n$-string over $\mc{X}$ by $x_k^{n+k-1} = x_kx_{k+1}\ldots x_{n+k-1} \in \mc{X}^{n}$. Further, let $x^n = x_1^n$ and for $i>n$, $x_i^n = \emptyset$, where $\emptyset$ denotes the null string.
Let ${\mu^n}$ denote a multinomial probability distribution on the sample space $\mc{X}^n$, where $\mu^1$ is the corresponding categorical distribution on $\mc{X}$, i.e., $\mu^n(x^n) = \prod_{i=1}^n \mu^1(x_i)$.
%We refer to the sequence $\mub$ of the probability measures on $n$-strings as a string-source.
We refer to $\mub:= \{ \mu^n : n \in \mathbb{N}\}$ as a (memoryless) string-source, also called source.
%Let $\mc{M}_{\mc{X}}$ denote the set of all string-sources on the sample space $\mc{X}$.
% A string-source is said to be stationary if $\sum_{x_1,\ldots,x_k}\mu^{n+k}(x^{n+k})  = \mu^n(x_{k+1}^n).$ 
Let $\theta= (\theta_1, \ldots, \theta_{|\mc{X}|})$ be the parameter vector corresponding to the categorical distribution $\mu^1$. That is 
$\theta_i : = \mu^1(a_i) = P[X = a_i] $ for all $i \in [|\mc{X}|]$. 
%Denote ${\mu_\theta^n} $ as the product distribution of $\mu^1_\theta$ on $\mc{X}^n$, i.e., .
By definition, $\theta$ is an element of the $(|\mc{X}|-1)$ dimensional simplex of all stochastic vectors of size $|\mc{X}|$. Denote $\Theta_{|\mc{X}|}$ as the $|\mc{X}|$-ary probability simplex, i.e.,
$
\Theta_{|\mc{X}|} : = \left\{ \theta \in \mathbb{R}_{+*}^{|\mc{X}|} : \sum_{i \in [|\mc{X}|]} \theta_i = 1\right\}.
$
We typically reserve upper case letters for random variables while  lower case letters correspond to their realizations. For example, $X^n$ denotes an $n$-string randomly drawn from $\mub$ while $x^n \in \mc{X}^n$ is a realization of the random $n$-string.

Note that as string lengths become large, the logarithm of the guesswork normalized by string length has been shown
 to satisfy a LDP for a more general set of sources~\cite{duffy-LDP}. 
 While existing results on guesswork consider the ordering of strings from most likely to least likely, either unconditionally or conditionally (starting at \cite{Arikan-guesswork}), or universal guessing orders \cite{sundaresan-universal}, the more restrictive assumptions used here enable us to treat, for the first time, the reverse guesswork process. In reverse guessing, strings are ordered from least likely to most likely, breaking ties arbitrarily. At first glance, consideration of the worst possible strategy seems to be of limited value, but we will show that is not the case. 

Forward guesswork provides good estimates of the likelihood of quickly discovered strings, but we shall demonstrate reverse guesswork provides better estimates on the likelihood that large numbers of guesses occur. We also establish an inherent geometrical correspondence between string-sources whose orders are reversed.  Amongst other results, we prove that logarithm of reverse guesswork scaled by string length satisfies a LDP with a concave rate function.  The concavity captures the fact that as one moves along the reverse order the rate of acquisition of probability increases initially.  Due
 to that fact, 
the approach employed in~\cite{duffy-LDP}, which leveraged earlier results on the scaling behavior of moments of guesswork, has no hope of success and we instead use distinct means.

\begin{definition}[information]
Let information $\imath_{\mub}^n: \mc{X}^n \to \mathbb{R}_{+*}$ be defined as:\footnote{$\mathbb{R}_{+*}:= \{ x\in \mathbb{R} | x>0\}$.}
\begin{equation}
\imath_{\mub}^n(x^n) := \log \frac{1}{\mu^n(x^n)}.
\label{eq:def-information}
\end{equation}
\end{definition}
Note that throughout this paper we use $\log(\cdot)$ to denote the natural logarithm and hence  information is measured in nats as defined above.

\begin{definition}[Shannon entropy]
Let $H^n(\mub)$ denote the entropy, which is defined as the expected information of a random $n$-string, and quantifies the average uncertainty that is revealed when a random $n$-string drawn from $\mub$ is observed:
%\footnote{In this paper $\log(\cdot)$ always denotes the logarithm in base $2$.} 
\begin{align}
H^n(\mub) := E_{\mub}\{ \imath_{\mub}^n(X^n)\}  =\sum_{x^n\in \mc{X}^n}{\mu^n}(x^n)\imath^n_\mub(x^n),
\label{eq:entropy}
\end{align}
where $E_{\mub}\{ \cdot\}$ denotes the expectation operator with respect to the measure $\mub$. 
The entropy rate, is denoted by $H(\mub)$ and is given by
$
H(\mub)  :=  \lim_{n\to \infty} \frac{1}{n} H^n(\mub) = H^1(\mub).
%\label{eq:entropy_rate}
$
\end{definition}

\begin{definition}
For any $n\in \mathbb{N}$, we define $[n] := \{1,\ldots,n\}$.
\label{def-[n]}
\end{definition}

In what follows we put forth two assumptions on the string-source that enable us to prove that the logarithm of reverse guesswork satisfies a LDP. We will also later characterize the set of string-sources that satisfy the assumptions. 
\begin{assumption}
We assume that $\mu^1(x)>0$ for all ${x \in \mc{X}}$, i.e., $\theta_i>0$ for all $i \in [|\mc{X}|]$.
\label{assump-rate}
\end{assumption}

Define
\begin{align}
\underline{\varepsilon}: = \min_{x \in \mc{X}}  \mu^1(x) \label{eq:max-min-prob},\\
\overline{\varepsilon}: = \max_{x \in \mc{X}}  \mu^1(x) \label{eq:min-max-prob}.
\end{align}
Assumption~\ref{assump-rate} excludes the boundaries of the probability simplex and ensures that the likelihood of the least likely $n$-string scales exponentially with $n$. In other words, for all $n \in \mathbb{N}$, and all $x^n \in \mc{X}^n$
\begin{equation}
 \frac{1}{n} \log \frac{1}{\mu^n(x^n)} \leq \log \frac{1}{\underline{\varepsilon}} < \infty.
\end{equation}

\begin{assumption} 
We assume that $\arg \min_{x \in \mc{X}}  \mu^1(x)$ and $\arg \max_{x \in \mc{X}}  \mu^1(x)$ are unique.
\label{assump-zero-entropy}
\end{assumption}
Assumption~\ref{assump-zero-entropy} ensures that the most likely and the least likely words are unambiguously determined. 

 The probability simplex for a ternary alphabet, $\Theta_{3}$, is depicted in Fig.~\ref{fig:trajectory}, where the green dashed and green solid lines represent the set of string-sources in $\Theta_3$ that do not satisfy Assumption~\ref{assump-zero-entropy}. Note that in this case, $(|\mc{X}|-2) = 1$ and ${{|\mc{X}|}\choose{2}}  =3$ resulting in three $1$-dimensional hyperplanes.

\begin{definition}[set $\mc{M}_{\mc{X}}$ of string-sources]
We denote by $\mc{M}_{\mc{X}}$, the set of all memoryless string-sources on alphabet $\mc{X}$ that satisfy Assumptions~\ref{assump-rate} and~\ref{assump-zero-entropy}.
\end{definition}
In this paper, we focus our attention to string-sources in $\mc{M}_{\mc{X}}$.

\begin{definition}[guesswork/optimal ordering for $\mub$]
A one-to-one function $G^n_\mub:  \mc{X}^n \to [|\mc{X}|^n]$ is said to be the optimal ordering for $\mub$ (on $n$-strings) if
\begin{equation}
\mu^n(x^n) > \mu^n(y^n) \Rightarrow G_\mub^n(x^n) < G_\mub^n(y^n),
\label{eq:optimal_ordering}
\end{equation}
and further if $\mu^n(x^n) = \mu^n(y^n)$, then $G_\mub^n(x^n)<G_\mub^n(y^n)$ if and only if $x^n$ appears prior to $y^n$ in lexicographic ordering. 
\end{definition}

\begin{definition}[reverse guesswork/reverse ordering for $\mub$]
We define $R_\mub^n(x^n)$ as the reverse guesswork, given by
\begin{equation}
R_\mub^n(x^n) := |\mc{X}|^n - G^n_{\mub}(x^n)
\label{eq:reverse-guesswork}
\end{equation}
\end{definition}

{Note that as defined above, given $\mub$, guesswork and reverse guesswork are unique functions.}
While guesswork corresponds to the best guessing strategy for recovering the string $x^n$, reverse guesswork corresponds to the worst guessing strategy for the string-source $\mub$ as it proceeds from the least likely string all the way to the most likely. Hence, the statistical performance of any other guessing strategy (ordering on $n$-strings) lies somewhere between that of guesswork and reverse guesswork. As will become clear later in the paper, our interest in reverse guesswork is mainly due to its usefulness in characterizing the PMF of guesswork.

\begin{definition}[logarithm of forward/reverse guesswork]
We define the logarithm of (forward) guesswork $g_\mub^n: \mc{X}^n \to [|\mc{X}|^n]$ and the logarithm of reverse guesswork $r_\mub^n: \mc{X}^n \to [|\mc{X}|^n]$ as
\begin{align}
g_\mub^n(x^n) &:= \log G^n_{\mub}(x^n),\\
r_\mub^n(x^n) &:= \log R^n_{\mub}(x^n),
\end{align}
where $G^n_\mub$ and $R^n_\mub$ are guesswork and the reverse guesswork defined in~\eqref{eq:optimal_ordering} and~\eqref{eq:reverse-guesswork}, respectively. 
\label{def:exponent}
\end{definition}

\begin{definition}[varentropy~\cite{strassen}]
Let the varentropy, denoted by $V^{n}(\mub)$, be defined as
\begin{align}
V^{n}(\mub) &= \var_{\mub} \left\{\imath_{\mub}^n(X^n)\right\} 
\label{eq:varentropy}
\end{align}
where $\var_{\mub}\{ \cdot\}$ denotes the variance operator with respect to the measure $\mub$.
The varentropy rate is denoted by $V(\mub)$ and is given by
\begin{equation}
V(\mub) := \lim_{n \to \infty} \frac{1}{n} V^{n}(\mub) = V^1(\mub).
%\vspace{-.15in}
\label{eq:varentropyrate}
\end{equation}
\end{definition}

\begin{definition}[relative entropy]
Denote $D^n(\mub||\rhob) $ as the relative entropy, also referred to as the Kullback-Leibler (KL) divergence, between the measures $\mu^n$ and $\rho^n$, given by
\begin{equation}
D^n(\mub||\rhob) := \sum_{x^n\in \mc{X}^n}{\mu^n}(x^n) \log\left(\frac{{\mu^n}(x^n)}{{\rho^n}(x^n)}\right) = nD^1(\mub||\rhob).
\end{equation}
\end{definition}

\begin{definition}[cross entropy]
Let the cross entropy, denoted by $H^n(\mub||\rhob)$, be defined as
\begin{align}
H^n(\mub||\rhob) := &E_\mub \left\{\log\left(\frac{1}{{\rho^n}(X^n)}\right)\right\}\label{eq:HH}\\
 = &\sum_{x^n\in \mc{X}^n}{\mu^n}(x^n) \log\left(\frac{1}{{\rho^n}(x^n)}\right).
\nonumber\\
= &  H^n(\mub) + D^n(\mub||\rhob) \label{eq:HHH}\\
= & nH^1(\mub) + n D^1(\mub||\rhob).\nonumber
\end{align}
Observe that $H^n(\mub|| \mub)= H^n(\mub)$ recovers the Shannon entropy.
\end{definition}

As a natural generalization of  varentropy and cross entropy, we define cross varentropy in the following definition.
\begin{definition}[cross varentropy]
Let the cross varentropy, denoted by $V^n(\mub||\rhob)$, be defined as
\begin{align}
V^n(\mub||\rhob) := & \var_\mub \left\{\log\left(\frac{1}{{\rho^n}(X^n)}\right)\right\}\nonumber\\
 = &\sum_{x^n\in \mc{X}^n}{\mu^n}(x^n) \left[\log\left(\frac{1}{{\rho^n}(x^n)}\right) - H^n(\mub||\rhob) \right]^2. 
\label{eq:VV}
\end{align}
Note that $V^n(\mub|| \mub)= V^n(\mub)$ simplifies to the varentropy. Further note that  $V^n(\mub||\rhob)=  n V^1(\mub||\rhob)$ because of our restriction to i.i.d. string-sources.
\label{def:mismateched-varentropy}
\end{definition}

An operational interpretation of the cross entropy $H^n(\mub || \rhob)$ arises in prefix-free source coding, where it is known that, ignoring the integer constraints on the codeword length, the length of the codeword for the string $x^n$ that minimizes the average codeword length equals $\imath^n_\mub(x^n) = - \log \mu^n(x^n)$. In this case, the resulting average codeword length is given by
\begin{equation}
E_\mub \{ \imath^n_\mub(X^n)\} =  E_\mub \left\{ \log \frac{1}{\mu^n(X^n)}\right\} = H^n(\mub).
\end{equation}
Therefore, if a source code is designed for the mismatched string-source $\rhob$, the length of the codeword for $x^n$ would be chosen to be $\imath_\rhob^n(x^n)$. Hence, the average codeword length of the optimal prefix-free source code that is designed to minimize the average codeword length under the (mismatched) distribution $\rhob$ is equal to the cross entropy:
\begin{equation}
E_\mub \{ \imath^n_\rhob(X^n)\} = E_\mub \left\{ \log \frac{1}{\rho^n(X^n)}\right\} = H^n(\mub||\rhob).
\end{equation}

\begin{definition}[R\'enyi entropy]
For any distribution $\mu^n$, let the R\'enyi entropy of order $\alpha$, denoted by $H^n_{\alpha}(\cdot)$, be defined as~\cite{Renyi-entropy}
\begin{equation}
H^n_\alpha(\mub) := \frac{1}{1-\alpha} \log \left(\sum_{x^n \in \mc{X}^n} [\mu^n(x^n)]^\alpha \right).
\label{eq:renyi_entropy}
\end{equation}
Further, define the R\'enyi entropy rate of order $\alpha$ (if it exists) as
\begin{equation}
H_\alpha(\mub) := \lim_{n \to \infty} \frac{1}{n}H^n_\alpha(\mub) = H^1_\alpha(\mub).
\end{equation}
Note that $\lim_{\alpha \to 1}H_\alpha(\mub) = H(\mub),$ which recovers the Shannon entropy.
\label{def:renyi_entropy}
\end{definition}

\subsection{The tilt}

The primary objective of this paper is to develop a tool, called the tilt, for the characterization of guesswork, its moments, and its large deviations behavior.
The tilt operation $T$ on any string-source $\mub$ is formally defined below.
\begin{definition}[tilted string-source of order $\alpha$]
Let the map $T$ operate on a string-source $\mub$ and a real-valued parameter $\alpha \in \mathbb{R}$ and output a string-source called ``tilted $\mub$ of order $\alpha$'', which is denoted by $T(\mub, \alpha)$ and is defined as
\begin{equation}
T(\mub, \alpha ) := \{ T(\mu^n, \alpha) : n \in \mathbb{N}\},
\end{equation}
where $T(\mu^n, \alpha)$ with overloading the notation
for all $n \in \mathbb{N}$ and all $x^n\in \mc{X}^n$ is defined as
\begin{equation}
T(\mu^n, \alpha) (x^n) : = \frac{[\mu^n(x^n)]^\alpha}{\sum_{y^n \in \mc{X}^n} [\mu^n(y^n)]^\alpha}.
\label{eq:tilt}
\end{equation}
\end{definition}
Note that the tilt is a real analytic function of $\alpha$ in $\mathbb{R}$ for all $\mub \in \mc{M}_\mc{X}$ and all $n \in \mathbb{N}$, formally stated in Lemma~\ref{lem:properties}.
Tilted $\mub$ of order $1$ is the string-source $\mub$ itself, i.e., $T(\mub,1) = \mub$. Further, for any $\mub \in \mc{M}_{\mc{X}}$, $T(\mub, 0) = \mb{u}_{\mc{X}}$, where $\mb{u}_{\mc{X}}$ is defined as the uniform string-source on alphabet $\mc{X}$ such that for all $x^n \in \mc{X}^n$
\begin{equation}
u_{\mc{X}}^n(x^n) := \frac{1}{|\mc{X}|^n}.
\label{eq:def-uniform}
\end{equation}
Notice that by definition $\mb{u}_\mc{X} \not \in \mc{M}_\mc{X}.$

Observe that  for all $n \in \mathbb{N}$, $T(\mu^n, \alpha)$ is a multinomial distribution with the corresponding categorical distribution given by $\tau(\theta,\alpha) = (\tau_1(\theta,\alpha), \ldots, \tau_{|\mc{X}|} (\theta,\alpha))$, where $\tau_i: \Theta_{|\mc{X}|} \times \mathbb{R} \to \Theta_{|\mc{X}|}$ for all $i \in [|\mc{X}|]$ is given by
\begin{equation}
\tau_i(\theta,\alpha) := \frac{\theta_i^\alpha}{\sum_{\i=1}^{|\mc{X}|} \theta_i^\alpha}.
\end{equation}
Let $\Gamma^{+}_\theta \in \Theta_{|\mc{X}|}$ denote the ``tilted family of $\theta$'' and be given by
\begin{equation}
\Gamma^{+}_\theta:= \{\tau(\theta,\alpha) :{\alpha \in \mathbb{R}_{+*}}\}.
\end{equation}
Observe that $\Gamma^{+}_\theta \in \Theta_{|\mc{X}|}$ is a set of stochastic vectors in the probability simplex.
Thus, the tilted family of a memoryless string-source with parameter vector $\theta$ is comprised of a set of memoryless string-sources whose parameter vectors belong to the tilted family of the vector $\theta$, i.e., $\Gamma^+_\theta$.

\begin{definition}[reversed string-source]
We define the reverse of the string-source $\mub$ (also called $\mub$-reverse), denoted by $T ( \mub, -1).$
\end{definition}
Note that the guesswork corresponding to $\mub$-reverse is exactly the reverse guesswork. 
As an example, the reverse of a binary memoryless string-source with parameter vector $\theta = (p, 1-p)$ is a binary memoryless string-source with $\tau(\theta, -1) = (1-p, p)$. Observe that $R_\mub^n$ is the optimal ordering (guesswork) for the reverse string source $T ( \mub, -1).$ Further, $G_\mub^n$ is the reverse ordering for $T ( \mub, -1).$

\begin{definition}[tilted/reverse family of a string-source]
Let the set $\mc{T}^+ _{\mub}$ denote the tilted family of the string-source $\mub$, and be given by
\begin{equation}
\mc{T}^+ _{\mub}:= \{  T(\mub,\alpha) : \alpha \in \mathbb{R}_{+*}\}.
\end{equation}
Further, define the reverse family of the string-source as
\begin{equation}
\mc{T}_{\mub}^-:=\{ T( \mub, \alpha): \alpha\in \mathbb{R}_{-*}\}.
\end{equation}
\vspace{0.05in}
\end{definition}
The first notable property of the tilted family of $\mub$ is that it parametrizes a standard exponential family of string-sources with $\alpha$. In other words, for any $n \in \mathbb{N}$, the family of distributions $\{T( \mu^n, \alpha): \alpha \in \mathbb{R}\}$ parametrized by $\alpha$ characterizes an exponential family of distributions on $\mc{X}^n$.

\begin{figure*}
\centering
\includegraphics[width=0.7\textwidth]{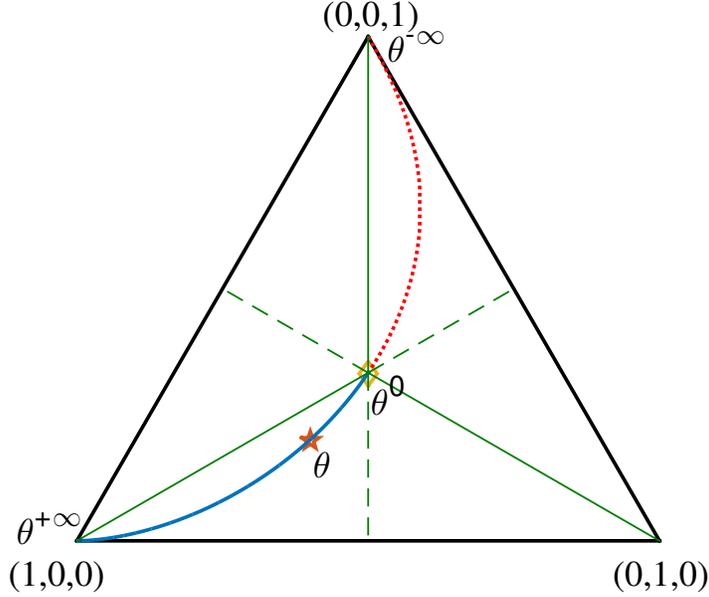}
\caption{The solid black lines depict the boundaries of the probability simplex $\Theta_{3}$ of all ternary stochastic vectors, which are excluded from our analysis as they do not satisfy Assumption~\ref{assump-rate}. 
The dashed green lines represent the string-sources that do not satisfy~\eqref{eq:entropy-00} in Assumption~\ref{assump-zero-entropy};
and the solid green lines represent the string-sources that do not satisfy~\eqref{eq:entropy-0} in Assumption~\ref{assump-zero-entropy}.
The red pentagon marker is  the ternary source parameter vector $\theta = (0.2, 0.3, 0.5)$; 
the blue solid curve depicts $\Gamma^+_\theta$, i.e., the tilted family of $\theta$;
the red dotted curve depicts $\Gamma^-_\theta$, i.e., the reverse family of $\theta$; 
the yellow diamond marker corresponds to the uniform parameter vector $\theta^0 = (\frac{1}{3},\frac{1}{3},\frac{1}{3} )$; and $\theta^\infty = (1,0,0)$ and $\theta^{-\infty} = (0,0,1)$. Note that $\theta^{\infty}$ and $\theta^{-\infty}$ are only achieved in the limit and do not belong to either $\Gamma^+_\theta$ or $\Gamma^-_\theta$.}
\label{fig:trajectory}
\end{figure*}

As a working example throughout the paper, we consider a memoryless string-source on a ternary alphabet ($|\mc{X}| = 3$) with the parameter vector $\theta = (0.2, 0.3, 0.5)$. The ternary alphabet is general enough to encompass the difficulties of the problem while simple enough for us to demonstrate the geometric techniques developed in this paper.
The ternary probability simplex is depicted in Fig.~\ref{fig:trajectory}, where the tilted family 
$\Gamma^+_\theta$ is plotted for $\theta = (0.2, 0.3, 0.5)$. As can be seen, the tilted family of $\theta$ is a parametric curve that is contained within the 2-dimensional simplex of stochastic vectors. 
The curve starts arbitrarily close to the maximum entropy uniform parameter vector $\theta^0$ (for $\alpha = 0^+$) and then moves towards one of the zero-entropy corner points of the simplex corresponding to the most likely symbol as $\alpha \to \infty$.
We shall show that these properties hold generally for all string-sources in $\mc{M}_{\mc{X}}$.

As a final note, the tilt is intimately related to the  concept of information geometry (see~\cite{information-geometry}), where several geometric properties of the Kullback-Leibler divergence (relative entropy) have been studied and their relevance to problems in information theory have been investigated. In particular, Borade and Zheng applied concepts from information geometry to the analysis of error exponents~\cite{lizhong-allerton-2006}; and Huang {\em et al.} performed a geometric study of hypothesis testing~\cite{meyn-entropy}.
This work also builds upon the insights on the geometry of the probability simplex in the context of information theory and statistics pioneered by Barron, Clarke, and Cover~\cite{Barron_Cover_91,Clarke_Barron}. 
The preliminary version of this work relied on the method of types (see~\cite{Allerton15_guesswork}) while in this paper we build on a  generalized notion of the weakly typical set, called the tilted weakly typical sets. We believe that the current treatment may be generalized to a  class of stationary ergodic string-sources as discussed in the concluding remarks.

\subsection{Organization and contribution}

In Section~\ref{sec:properties}, we provide the main properties of the tilt operation. In particular, we characterize the entropy $H^n(T(\mub, \alpha))$, the cross entropy $H^n(T(\mub, \alpha)||\mub)$, the relative entropy $D^n(T(\mub, \alpha)||\mub)$ of the tilited string-source as well as the R\'enyi entropy $H_\alpha^n(\mub)$. These quantities are used in Section~\ref{sec:LDP} to implicitly characterize the large deviations rate function of the logarithm of reverse guesswork, the logarithm of (forward) guesswork, and information. We further show that the derivatives (with respect to the parameter $\alpha$) of these quantities can be expressed in terms of the varentropy $V^n(T(\mub, \alpha))$ and the  cross varentropy $V^n(T(\mub, \alpha)||\mub)$ of the tilted string-source.

 In Section~\ref{sec:order_class}, we define the equivalent order class of a string-source as the set of all string-sources that result in the same guesswork for all $n$, and denote this set of string-sources by $\mc{C}_\mub$. In Theorem~\ref{thm:order_class}, we show that
the equivalent order class of $\mub$ is equal to the tilted family, i.e.,  ${\mc{C}_{\mub}} = \mc{T}^{+}_{\mub},$ providing an operational meaning to the tilted family.
Further, the set of all string-sources whose optimal orderings are inverse for $\mub$ is equal to $\mc{T}^-_{\mub}$, i.e., $\mc{C}_{T ( \mub, -1)} = \mc{T}^-_{\mub}$.
This result draws an important distinction between information and guesswork. The string-source $\mub$ is uniquely determined if information $\imath^n_\mub(x^n)$ is provided for all $x^n \in \mc{X}^n$. On the other hand, it is not possible to  recover uniquely the string-source $\mub$ if one is provided with the guesswork $G_\mub^n(x^n)$ for all strings $x^n \in \mc{X}^n$. Our result shows that one can only determine the string-source up to equivalence within the tilted family that the source belongs to given the guesswork for all $x^n \in \mc{X}^n$ leading to one less degree of freedom. This result has implications in terms of universal source coding~\cite{IT_Kosut_one2one, ITW14} and universal guesswork~\cite{sundaresan-universal, ISIT15_guesswork}, where the degree of freedom in the model class determines the penalty of universality, which is called redundancy in source coding.

In Section~\ref{sec:typical-set}, we generalize the notion of the weakly typical set. The weakly typical set comprises of all strings $x^n \in \mc{X}^n$ such that $ e^{-H^n(\mub) - n \epsilon}<\mu^n(x^n) < e^{-H^n(\mub) + n \epsilon}$. An important implication of the weakly typical set (known as the asymptotic equipartition property) is that the typical set roughly contains $\approx e^{H^n(\mub)}$ strings each of probability  $\approx e^{-H^n(\mub)}$. 
For sufficiently large $n$, the probability of the typical set is close to one, i.e., if a string is randomly drawn from the source it is in the weakly typical set with high probability. The weakly typical set is used to prove key results in information theory that are dominated by the typical events, such as the source coding theorem, and the channel coding theorem. Despite the great intuition that typicality provides on such quantities, the arguments based on typicality fall short in providing intuition on other key quantities that are dominated by atypical events, such as the error exponents in decoding, and the cut-off rate in sequential decoding. %What we provide is a natural ordering of all strings based on the tilt operation. 
In Section~\ref{sec:typical-set}, we consider all strings $x^n \in \mc{X}^n$ such that $e^{-H^n(T(\mub, \alpha)||\mub) - n \epsilon}<\mu^n(x^n) < e^{-H^n(T(\mub, \alpha)||\mub) + n \epsilon}$ for $\alpha \in \mathbb{R}$, which we define as the tilted weakly typical set of order $\alpha$ and denote by $\mc{A}^n_{\mub,\alpha,\epsilon}$ (see Definition~\ref{def:typical-set}). We show that for a fixed $\epsilon$, for any string $x^n \in \mc{X}^n$, there exists an $\alpha$ such that $x^n$ is contained in the tilted weakly typical set of order $\alpha$. Hence, this notion could be used to cluster/partition all of the strings $x^n \in \mc{X}^n$ based on their likelihoods.
In Theorem~\ref{thm:LDP-information}, we derive tight bounds on the probability, size, guesswork, and reverse guesswork for all elements of the tilted weakly typical set of any order $\alpha$.
Roughly speaking, for any $\alpha \in \mathbb{R}_{+*}$, we show that $\mc{A}^n_{\mub,\alpha,\epsilon}$ contains $\approx e^{H^n(T(\mub, \alpha))}$ strings where the likelihood of each string is $\approx e^{-H^n(T(\mub, \alpha)||\mub)}$, and the probability of the entire set is $\approx e^{-D^n(T(\mub, \alpha)||\mub)}$, and the guesswork for each element in the set is $\approx e^{H^n(T(\mub, \alpha))}$.
For any $\alpha \in \mathbb{R}_{-*}$, we show that $\mc{A}^n_{\mub,\alpha,\epsilon}$ contains $\approx e^{H^n(T(\mub, \alpha))}$ strings where the likelihood of each string is $\approx e^{-H^n(T(\mub, \alpha)||\mub)}$, and the probability of the entire set is $\approx e^{-D^n(T(\mub, \alpha)||\mub)}$, and the ``reverse'' guesswork for each element in the set is $\approx e^{H^n(T(\mub, \alpha))}$.
 Note that $\alpha =1$ recovers the weakly typical set as $H^n(T(\mub, 1)||\mub) = H^n(T(\mub, 1)) = H^n(\mub)$, and for $\alpha \neq 1$, this provides a generalization of the well-known  weakly typical set.
This generalization could be used to provide new insights on understanding and proving results on information theoretic quantities that are dominated by atypical events.

In Section~\ref{sec:LDP}, we prove that the logarithm of reverse guesswork $r_\mub^n(x^n)$, the logarithm of (forward) guesswork  $g_\mub^n(x^n)$, and information $\imath_\mub^n(x^n)$ satisfy a LDP and provide an implicit formula for the rate function in each case based on entropy, cross entropy, and relative entropy. Although some of the results in this section have been previously proved in more generality, we believe that the machinery built upon the tilt operation in this paper as well as the information theoretic characterization of the rate functions provide useful intuition in characterizing the large deviations behavior of these quantities. In particular, the LDP for the logarithm of reverse guesswork could not have been established using the previous methods by Christiansen and Duffy~\cite{duffy-LDP}, as the resulting rate function here is concave. 
The most common way to prove a LDP is the G{\"a}rtner-Ellis Theorem \cite[Theorem 2.3.6]{dembo-book}. There, one first determines the scaled cumulant generating function of the process of interest, which identifies how moments scale, and then leverages properties of it to establish the LDP. This approach can only work if the resulting rate function is convex. Here, that is not the case.

In Section~\ref{sec:approximation}, we provide an approximation on the distribution of guesswork that captures the behavior even for small string lengths. The approximation is built based on the tilt operation, guesswork, and reverse guesswork. We empirically show that the approximation can be applied to a variety of string-sources to characterize the PMF of guesswork.

Finally, the conclusion is provided in Section~\ref{sec:conclusion}, where we also present ideas on how to generalize these results beyond the setting considered in this paper.

\section{Properties of the Tilted String-Sources}
\label{sec:properties}
In this section, we provide the main properties of the tilt operation that shall be used in proving our results on the characterization of guesswork. This section may be skipped by the reader and only referred to whenever a particular lemma is needed.

%We start with two basic properties of the tilt operation.
\begin{lemma}
For any $\alpha, \beta \in \mathbb{R}$,
\begin{equation}
T\left(T(\mub, \alpha), \beta \right) =  T\left(T(\mub, \beta), \alpha \right) = T(\mub, \alpha \beta).
\label{eq:tiltoftilt}
\end{equation}
\label{lem:tiltoftilt}
\end{lemma}
\begin{IEEEproof}
This is obtained from the definition using algebraic manipulation.
\end{IEEEproof}

\begin{lemma}
The optimal ordering for $\mub$ is the reverse ordering for $T ( \mub, -1)$ and the reverse ordering for $\mub$ is the optimal ordering for $T ( \mub, -1)$.
\label{lemma:reverse}
\end{lemma}
\begin{IEEEproof}
The statement is a consequence of the definition.
\end{IEEEproof}

\begin{lemma}
For any $\mub \in \mc{M}_{\mc{X}}$,
\begin{align}
~&\lim_{\alpha \to \infty} H(T( \mub, \alpha)) = 0,\label{eq:entropy-00}\\
~&\lim_{\alpha \to -\infty} H(T( \mub, \alpha)) = 0.
\label{eq:entropy-0}
\end{align}
\end{lemma}
\begin{IEEEproof}
The Lemma is a direct consequence of Assumption~\ref{assump-zero-entropy}.
\end{IEEEproof}

Next, we provide a lemma that relates R\'enyi entropy, Shannon entropy, relative entropy, and cross entropy.
\begin{lemma}
For any $\mub \in \mc{M}_{\mc{X}}$, and any $\alpha \in \mathbb{R}$, the following relations hold:
\begin{equation}
H^n(T(\mub, \alpha)||\mub)  - H^n_\alpha(\mub)   = \frac{1}{1-\alpha} D^n(T(\mub, \alpha)||\mub),
\label{eq:renyi-eq1}
\end{equation}
\begin{equation}
H^n(T(\mub, \alpha)) - H^n_\alpha(\mub) = \frac{\alpha}{1-\alpha} D^n(T(\mub, \alpha)||\mub),
\label{eq:renyi-eq2}
\end{equation}
where $\left. \frac{1}{1-\alpha} D^n(T(\mub, \alpha)||\mub)\right|_{\alpha = 1}:= 0.$
\vspace{0.08in}\label{lem:renyi-eq}
\end{lemma}
\begin{IEEEproof}
To establish~\eqref{eq:renyi-eq1}, 
\begin{align}
&H^n(T(\mub, \alpha)||\mub)  - H^n_\alpha(\mub) -\frac{1}{1-\alpha} D^n(T(\mub, \alpha)||\mub) \nonumber\\
= & \sum_{x^n \in \mc{X}^n} T(\mu^n, \alpha)(x^n) \log \frac{1}{\mu^n(x^n)}\label{eq:ttt3}\\
&- \frac{1}{1-\alpha}\sum_{x^n \in \mc{X}^n} [\mu^n(x^n)]^\alpha\label{eq:ttt4}\\
& - \frac{1}{1-\alpha} \sum_{x^n \in \mc{X}^n} T(\mu^n, \alpha)(x^n) \log\left(\frac{[\mu^n(x^n)]^\alpha / \sum_{y^n \in \mc{X}^n} [\mu^n(y^n)]^\alpha}{\mu^n(x^n)} \right)\label{eq:ttt1}\\
&= 0,\label{eq:ttt2}
\end{align}
where~\eqref{eq:ttt1} follows from the definition, and~\eqref{eq:ttt2} holds because the term $\sum_{y^n \in \mc{X}^n} [\mu^n(y^n)]^\alpha$  in~\eqref{eq:ttt1} is a constant, and hence~\eqref{eq:ttt1} could be decomposed into two terms one of which cancels the right hand side of~\eqref{eq:ttt3}, and the other cancels~\eqref{eq:ttt4}. Finally,~\eqref{eq:renyi-eq2} is established by putting together~\eqref{eq:HHH} and~\eqref{eq:renyi-eq1}.
\end{IEEEproof}

\begin{lemma}
For any $\mub \in \mc{M}_{\mc{X}}$, and any $\alpha \in \mathbb{R}$, the following holds:
\begin{align}
&\log \frac{1}{T(\mu^n, \alpha)(x^n)} -H^n(T(\mub,\alpha)) \nonumber\\
&=\alpha\left(
\log \frac{1}{\mu^n(x^n)} - H^n(T(\mub,\alpha)||\mub) \right).
\end{align}
\label{lem:tilt-mean}
\end{lemma}
\begin{IEEEproof}
The lemma is proved by considering the following chain of equations:
\begin{align*}
&\log \frac{1}{T(\mu^n, \alpha)(x^n)} -H^n(T(\mub,\alpha))\\
& = \alpha \log \frac{1}{\mu^n(x^n)}  + \log \left(\sum_{x^n \in \mc{X}^n} [\mu^n(x^n)]^\alpha\right)\\
&- \sum_{x^n \in \mc{X}^n} T(\mu^n,\alpha)(x^n) \left( \alpha \log \frac{1}{\mu^n(x^n)}  + \log \left(\sum_{x^n \in \mc{X}^n} [\mu^n(x^n)]^\alpha\right) \right)\\
& = \alpha \left(\log \frac{1}{\mu^n(x^n)} - H^n(T(\mub,\alpha)||\mub)\right).
\end{align*}
\end{IEEEproof}

\begin{lemma}
For any $\mub \in \mc{M}_{\mc{X}}$, and any $\alpha \in \mathbb{R}$, the following holds:
\begin{equation}
\alpha ^2 V^n(T(\mub, \alpha) || \mub) =    V^n(T(\mub, \alpha)).
\end{equation}
\label{lem:mismatched-varentropy}
\end{lemma}
\begin{IEEEproof}
We have
\begin{align}
&\alpha^2 V^n(T(\mub, \alpha) || \mub) \nonumber\\
& = \sum_{x^n\in \mc{X}^n}{T(\mu^n, \alpha)}(x^n) \left[\alpha \left(\log\left(\frac{1}{{\mu^n}(x^n)}\right) - H^n(T(\mub, \alpha)||\mub) \right)\right]^2\nonumber\\
& = \sum_{x^n\in \mc{X}^n}{T(\mu^n, \alpha)}(x^n) \left[\log \frac{1}{T(\mu^n, \alpha)(x^n)} -H^n(T(\mub,\alpha)) \right]^2\label{eq:zxc1}\\
& = V^n(T(\mub, \alpha)),\nonumber
\end{align}
where~\eqref{eq:zxc1} follows from Lemma~\ref{lem:tilt-mean}.
\end{IEEEproof}

\begin{lemma}
For any string-source $\mub \in \mc{M}_{\mc{X}}$, for any $n\in \mathbb{N}$, $T( \mu^n, \alpha)$ is a real analytic function of $\alpha \in \mathbb{R}$. In particular, $\frac{d}{d\alpha} T( \mu^n, \alpha)$ exists; and  
$
\lim_{\alpha \to 0} T( \mu^n, \alpha) = u_\mc{X}^n,
$
where $u_{\mc{X}}^n$ is the uniform distribution on $\mc{X}^n$.
%, then, $T(\theta,\alpha)$
\label{lem:properties}
\end{lemma}
\begin{IEEEproof}
The statement follows from the facts that the exponential function is real analytic in its argument and that the denominator in~\eqref{eq:tilt} in the definition of the tilt is non-zero for all $\alpha \in \mathbb{R}$.
The limit as $\alpha \to 0$ holds since $\mu^n(x^n)>0$ for all $x^n \in \mc{X}^n$ by Assumption~\ref{assump-rate}. %The last part of the claim was assumed in~\eqref{eq:entropy-0} in Assumption~\ref{assump-rate}.
\end{IEEEproof}

\begin{lemma}
For any string-source $\mub \in \mc{M}_{\mc{X}}$, for any $n\in \mathbb{N}$, $H^n(T( \mub, \alpha))$, $D^n(T( \mub, \alpha)||\mub)$,  $H^n(T( \mub, \alpha)||\mub)$,  $V^n(T( \mub, \alpha))$, and $V^n(T( \mub, \alpha)||\mub)$ are continuous and differentiable functions of $\alpha \in \mathbb{R}$.
\label{lem:properties2}
\end{lemma}
\begin{IEEEproof}
Note that $H^n(\rhob)$, $D^n(\rhob||\mub)$, $H^n(\rhob||\mub)$, $V^n(\rhob)$, and $V^n(\rhob||\mub)$ are continuously differentiable with respect to $\rhob \in \mc{M}_{\mc{X}}$, and by Lemma~\ref{lem:properties}, $T(\mub, \alpha)$ is also continuous and differentiable with respect to $\alpha \in \mathbb{R}$ , and hence the overall composition by considering $\rhob = T(\mub, \alpha)$ is continuous and differentiable with respect to $\alpha \in \mathbb{R}$.
\end{IEEEproof}

\begin{lemma}
For any $\mub \in \mc{M}_{\mc{X}}$, and any $n \in \mathbb{N}$, and any $\alpha \in \mathbb{R}$:
\begin{align}
&\frac{d}{d\alpha} T(\mu^n, \alpha)(x^n)\nonumber\\
 &= T(\mu^n, \alpha)(x^n) \left( H^n(T(\mub, \alpha)||\mub) -  \log \frac{1}{\mu^n(x^n)} \right).
\label{eq:derivative_tilt}
\end{align}
\label{lem:derivative_tilt}
\end{lemma}
\begin{IEEEproof}
Let 
\begin{equation*}
T'(\mu^n , \alpha_0)(x^n) := \left. \frac{d}{d\alpha}T(\mu^n, \alpha)(x^n)\right|_{\alpha = \alpha_0}.
\end{equation*}
Note that by Lemma~\ref{lem:tiltoftilt}, we know that for any $\beta >0$:
\begin{equation*}
T(\mu^n, \alpha)(x^n) = T\left(T(\mu^n, \beta), \frac{\alpha}{ \beta}\right)(x^n).
\end{equation*}
Differentiating both sides with respect to $\alpha$, for all $\alpha_0, \beta>0$ we  have
\begin{align}
\left. \frac{d}{d\alpha}T(\mu^n, \alpha)(x^n)\right|_{\alpha = \alpha_0} &= T'(\mu^n , \alpha_0)(x^n)\nonumber\\   &= \left. \frac{d}{d\alpha}T\left(T(\mu^n, \beta), \frac{\alpha}{ \beta}\right)(x^n)\right|_{\alpha = \alpha_0}\nonumber\\
&= \frac{1}{\beta} T'\left(T(\mu^n, \beta), \frac{\alpha_0}{ \beta}\right)(x^n)\nonumber\\
& =\frac{1}{\alpha_0} T'\left(T(\mu^n, \alpha_0), 1\right)(x^n)  \label{eq:beta-alpha0},
\end{align}
where~\eqref{eq:beta-alpha0} follows by setting $\beta = \alpha_0.$ In other words,~\eqref{eq:beta-alpha0} implies that the derivative of a tilt with respect to $\alpha$ can be derived by taking the derivative at $\alpha =1$ and rescaling the derivative evaluated at the tilted distribution.
Consequently, to prove the lemma it suffices to prove that
\begin{equation}
\left. \frac{d}{d\alpha} T(\mu^n, \alpha)(x^n)\right|_{\alpha = 1} = \mu^n(x^n) \left(  H^n(\mub) - \log \frac{1}{\mu^n(x^n)}  \right),
\label{eq:derivative_tilt_1}
\end{equation}
and then it follows from~\eqref{eq:beta-alpha0} that for any $\alpha \neq 0$, we will arrive at the following:
\begin{align*}
&\frac{d}{d\alpha} T(\mu^n, \alpha)(x^n)\nonumber\\
&=  \frac{1}{\alpha}T(\mu^n, \alpha)(x^n) \left( H^n(T(\mub, \alpha)) -  \log \frac{1}{T(\mu^n, \alpha)(x^n)} \right).
\end{align*}
On the other hand, the desired result is obtained by considering Lemma~\ref{lem:tilt-mean}, and noting the continuity and differentiability of $T(\mu^n, \alpha)$ at $0$ from Lemma~\ref{lem:properties} and that of $H(T(\mu^n, \alpha))$ at $0$ from Lemma~\ref{lem:properties2}.

To establish~\eqref{eq:derivative_tilt_1}, note the definition of the tilt in~\eqref{eq:tilt} and the fact that the derivative of the numerator with respect to $\alpha$ is $[\mu^n(x^n)]^\alpha \log \mu^n(x^n)$ and the derivative of the denominator with respect to $\alpha$ is -$H^n(\mub).$
\end{IEEEproof}

\begin{lemma}
For any $\mub \in \mc{M}_{\mc{X}}$, we have
\begin{equation}
\lim_{\alpha \to 0} \frac{1}{\alpha^2} V^n(T(\mub, \alpha))=  V^n( \mb{u}_\mc{X}|| \mub).
\end{equation}
\label{lem:alpha0}
\end{lemma}
\begin{IEEEproof}
Observe that
\begin{align}
\lim_{\alpha \to 0} \frac{1}{\alpha^2} V^n(T(\mub, \alpha))& = \lim_{\alpha \to 0} V^n(T(\mub, \alpha)||\mub) \label{eq:777}\\
& = V^n(\mb{u}_\mc{X} || \mub)\nonumber,
\end{align}
where~\eqref{eq:777} follows from Lemma~\ref{lem:mismatched-varentropy}, completing the proof.
\end{IEEEproof}

\begin{lemma}
For any $\mub \in \mc{M}_{\mc{X}}$, and any $\alpha \in \mathbb{R}$,
\begin{equation}
\frac{d }{d \alpha}H^n(T( \mub, \alpha)) = -\alpha V^n(T( \mub, \alpha)||\mub).
\end{equation}
\label{lem:H_decrease}
\end{lemma}
\begin{IEEEproof}
First, owing to Lemma~\ref{lem:properties2}, the derivative of $H^n(T( \mu^n, \alpha))$ with respect to $\alpha$ exists.

To prove the result for all $\alpha \neq 0$, it suffices to show that 
\begin{equation}
\left.\frac{d }{d \alpha}H^n(T( \mub, \alpha))\right|_{\alpha=1} = - V^n(\mub)
\label{eq:intermediate-step1}
\end{equation}
for any given $\mub \in \mc{M}_{\mc{X}}$. Then, by Lemma~\ref{lem:tiltoftilt}, $T(\cdot, \alpha +d\alpha) = T(T(\cdot, 1+ d\alpha/\alpha), \alpha)$ which for $\alpha \neq 0$ leads to (see~\eqref{eq:beta-alpha0} in the proof of Lemma~\ref{lem:derivative_tilt} for a more formal statement of this part of the claim):
\begin{equation*}
\frac{d }{d \alpha}H^n(T( \mub, \alpha)) = -\frac{1}{\alpha} V^n(T( \mub, \alpha)).
\end{equation*}
This, in conjunction with Lemma~\ref{lem:mismatched-varentropy} and continuity and differentiability of $H^n(T( \mub, \alpha))$ with respect to all $\alpha \in \mathbb{R}$ from Lemma~\ref{lem:properties2} leads to the desired result.

To show that~\eqref{eq:intermediate-step1} holds, we have
\begin{align}
&\hspace{-.1in}\left.\frac{d }{d \alpha}H^n(T( \mub, \alpha))\right|_{\alpha = 1} \nonumber\vspace{0.1in}\\
&= \sum_{x^n\in \mc{X}^n}\left.\frac{d}{d\alpha}  T( \mu^n, \alpha)(x^n) \right|_{\alpha=1} \log \frac{1}{\mu^n(x^n)}\nonumber\\
&~~~~~ + \sum_{x^n\in \mc{X}^n} \mu^n(x^n)  \left.\frac{d}{d\alpha} \log \frac{1}{ T( \mu^n, \alpha)(x^n)} \right|_{\alpha=1}   \nonumber\\
& = \sum_{x^n \in \mc{X}^n} \mu^n(x^n)\left(H^n(\mub) - \log \frac{1}{\mu^n(x^n)}  \right) \log\frac{1}{\mu^n(x^n)} \nonumber\\
& ~~~~~+ \sum_{x^n \in \mc{X}^n}\mu^n(x^n) \left(\log \frac{1}{\mu^n(x^n)}  - H^n(\mub) \right) \label{eq:dT}\\
& = E_{\mub}\left\{  \left(H^n(\mub) -\log \frac{1}{\mu^n(X^n)}\right) \log\frac{1}{\mu^n(X^n)} \right\}\nonumber\\
& = -\var_\mub \{ \imath_\mub^n(X^n)\}= - V^n(\mub),\label{eq:VE}
\end{align}
where~\eqref{eq:dT} follows from Lemma~\ref{lem:derivative_tilt}, and~\eqref{eq:VE} follows from the definition of the varentropy in~\eqref{eq:varentropy}. The limit as $\alpha \to 0$ also follow from Lemma~\ref{lem:alpha0}.
\end{IEEEproof}

\begin{lemma}
For any $\mub \in \mc{M}_{\mc{X}}$ and  any $\alpha \in \mathbb{R}$,
\begin{equation}
\frac{d }{d \alpha}H^n(T( \mub, \alpha)|| \mub) = - V^n(T(\mub, \alpha)||\mub).
\end{equation}
\label{lem:H_monotone}
\end{lemma}
\begin{IEEEproof}
We have
\begin{align}
&\left.\frac{d }{d \alpha}H^n(T( \mub, \alpha)||\mub)\right.\nonumber\\
&= \sum_{x^n\in \mc{X}^n}\left.\frac{d}{d\alpha}  T( \mu^n, \alpha)(x^n) \right. \log \frac{1}{\mu^n(x^n)}\nonumber\\
& = \sum_{x^n \in \mc{X}^n}T(\mu^n, \alpha)(x^n) \left(H^n(T(\mub, \alpha)||\mub) -\log \frac{1}{\mu^n(x^n)}\right) \log\frac{1}{\mu^n(x^n)} \label{eq:lem-diff}\\
& =  E_{T(\mub, \alpha)} \left\{\left(H^n(T(\mub, \alpha)||\mub) -\log \frac{1}{\mu^n(X^n)}\right) \log\frac{1}{\mu^n(X^n)}  \right\}\nonumber\\
& = - \var_{T( \mub, \alpha)} \{ \imath_{\mub}^n(X^n)\}\label{ineq1}\\
&= -V^n(T(\mub, \alpha)||\mub),\nonumber
\end{align}
where~\eqref{eq:lem-diff} follows from Lemma~\ref{lem:derivative_tilt}, and~\eqref{ineq1} follows from the definition of information in~\eqref{eq:def-information} and cross entropy in~\eqref{eq:HH} leading to the fact that
\begin{equation}
E_{T(\mub, \alpha)} \left\{\log \frac{1}{\mu^n(X^n)} \right\}  = H^n(T(\mub, \alpha)||\mub).
\end{equation}
\end{IEEEproof}

\begin{lemma}
For any $\mub \in \mc{M}_{\mc{X}}$ and  any $\alpha \in \mathbb{R}$,
\begin{equation}
\frac{d }{d \alpha}D^n(T( \mub, \alpha)|| \mub) = (\alpha-1)V^n(T( \mub, \alpha)||\mub).
\end{equation}
\label{lem:D_derivative}
\end{lemma}
\begin{IEEEproof}
Note that  from~\eqref{eq:HHH},
\begin{equation*}
D^n(T( \mub, \alpha)|| \mub) = H^n(T( \mub, \alpha)|| \mub) - H^n(T( \mub, \alpha)).
\end{equation*}
The lemma then follows by differentiating both sides with respect to $\alpha$ and applying Lemmas~\ref{lem:H_decrease} and~\ref{lem:H_monotone} on the right hand side.
\end{IEEEproof}

\begin{lemma}
For any $\mub \in \mc{M}_{\mc{X}}$ and  any $\alpha\neq1$,
\begin{equation}
\frac{d }{d \alpha}H_\alpha^n(\mub) = - \frac{1}{(1-\alpha)^2}D^n(T(\mub, \alpha)|| \mub),
\label{eq:cll1}
\end{equation}
and 
\begin{equation}
\left. \frac{d }{d \alpha}H_\alpha^n(\mub) \right|_{\alpha = 1} = - \frac{1}{2} V^n(\mub).
\label{eq:cll2}
\end{equation}
\label{lem:renyi_derivative}
\end{lemma}
\begin{IEEEproof}
To establish~\eqref{eq:cll1}, consider:
\begin{align}
\frac{d}{d\alpha}H^n_\alpha(\mub) &= \frac{d}{d\alpha}H^n(T(\mub, \alpha)|| \mub) - \frac{d}{d\alpha}\left(\frac{1}{1-\alpha} D^n(T(\mub, \alpha)|| \mub)\right)\label{eq:ooo1}\\
& = - \frac{1}{\alpha^2}V^n(T(\mub, \alpha)) + \frac{1}{\alpha^2}V^n(T(\mub, \alpha))\nonumber\\
& -  \frac{1}{(1-\alpha)^2}D^n(T(\mub, \alpha)|| \mub)\label{eq:ooo2}\\
& =- \frac{1}{(1-\alpha)^2}D^n(T(\mub, \alpha)|| \mub),\nonumber
\end{align}
where~\eqref{eq:ooo1} follows by invoking~\eqref{eq:renyi-eq1} in Lemma~\ref{lem:renyi-eq}, and~\eqref{eq:ooo2} follows from the chain rule of derivatives.

To establish~\eqref{eq:cll2}, we have
\begin{align}
\left. \frac{d }{d \alpha}H_\alpha^n(\mub) \right|_{\alpha = 1} &= \lim_{\alpha \to 1}\frac{d }{d \alpha}H_\alpha^n(\mub) \label{eq:d-cont}\\
& = \lim_{\alpha \to 1}  \frac{-D^n(T(\mub, \alpha)|| \mub)}{(1-\alpha)^2}\label{eq:cll3}\\
& =  \lim_{\alpha \to 1} - \frac{ \frac{\alpha -1}{\alpha^2 }V^n(T( \mub, \alpha))}{2 (\alpha - 1)}\label{eq:lopital1}\\
& = - \frac{1}{2} V^n(\mub),\nonumber
\end{align}
where~\eqref{eq:d-cont} follows from the continuity of the derivative of $H_\alpha^n(\mub)$ with respect to $\alpha$,~\eqref{eq:cll3} follows from~\eqref{eq:cll1}, and~\eqref{eq:lopital1} follows from L'H\^opital's rule and Lemma~\ref{lem:D_derivative}.\
\end{IEEEproof}
This result recovers the fact that R\'enyi entropies are monotonically decreasing with respect to the order, and further characterizes their rate of change with respect to the parameterization.

\begin{lemma}
For any $\mub \in \mc{M}_{\mc{X}}$,
\begin{align}
\lim_{n \to \infty} \frac{1}{n} V^n(T(\mub, \alpha)||\mub) & = V(T(\mub, \alpha)||\mub). \label{eq:lem-VV6}\\
\lim_{n \to \infty} \frac{1}{n} H^n(T(\mub, \alpha)) &=  H(T(\mub, \alpha)),\label{eq:lem-VV1}\\
\lim_{n \to \infty} \frac{1}{n}H^n(T(\mub, \alpha)||\mub) &=  H(T(\mub,\alpha)||\mub) ,\label{eq:lem-VV3}\\
\lim_{n \to \infty} \frac{1}{n} D^n(T(\mub, \alpha)||\mub) &=  D(T(\mub,\alpha)||\mub) ,\label{eq:lem-VV4}\\
\lim_{n \to \infty} \frac{1}{n} H_\alpha^n(\mub) &=  H_\alpha(\mub) ,\label{eq:lem-VV5}
\end{align}
Further, for all $\alpha \in \mathbb{R}$,
\begin{align}
\frac{d}{d\alpha} H(T(\mub, \alpha)) &= -\alpha V(T(\mub, \alpha)||\mub),\label{eq:lem-VVV1}\\
\frac{d}{d\alpha}  H(T(\mub,\alpha)||\mub)  & = - V(T(\mub, \alpha)||\mub),\label{eq:lem-VVV3}\\
\frac{d}{d\alpha}D(T(\mub,\alpha)||\mub) & = (\alpha-1)V(T(\mub, \alpha)||\mub).\label{eq:lem-VVV4}\\
\frac{d}{d\alpha}H_\alpha(\mub) & = - \frac{1}{(1-\alpha)^2}D(T(\mub, \alpha)|| \mub),\label{eq:lem-VVV5}
\end{align}
where $D(T(\mub, \alpha)|| \mub)$ is differentiable with respect to $\alpha$, and $\left.\frac{1}{(1-\alpha)^2}D(T(\mub, \alpha)|| \mub)\right|_{\alpha = 1}: = \frac{1}{2} V(\mub)$.
\label{lem:VV}
\end{lemma}
\begin{IEEEproof}
The first part of the claim is straightforward as the equalities hold for all $n$ and hence also hold in the limit as $n \to \infty.$
%Note that~\eqref{eq:lem-VV6} is a direct consequence of the fact for any $\mub \in \mc{M}_{\mc{X}}$ and $\alpha$, $V^n(\mub||T(\mub, \alpha)) = nV(\mub||T(\mub, \alpha))$. 
The claim in~\eqref{eq:lem-VVV1} hence follows   
by invoking Lemma~\ref{lem:H_decrease}. %Hence, the function $\frac{1}{n}H^n(T(\mub, \alpha))$ converges (as in~\eqref{eq:lem-VV1}) to a function as in that is differentiable with respect to $\alpha$ and its derivative is given by~\eqref{eq:lem-VVV1}. 
Equations~\eqref{eq:lem-VVV3} and~\eqref{eq:lem-VVV4} and~\eqref{eq:lem-VVV5} are established similarly considering Lemmas~\ref{lem:H_monotone},~\ref{lem:D_derivative}, and~\ref{lem:renyi_derivative}, respectively. 
\end{IEEEproof}

\begin{lemma}
If $\mub \in \mc{M}_{\mc{X}}$, then, for all $\alpha \neq 0$, we have $T(\mub,\alpha) \in \mc{M}_{\mc{X}}$.  In other words, $\mc{M}_{\mc{X}}$ is closed under all non-zero tilts, and hence $\mc{T}^+_\mub, \mc{T}^-_{\mub} \subseteq \mc{M}_{\mc{X}}$.
\label{lem:Theta2}
\end{lemma}
\begin{IEEEproof}
We need to show that $T(\mub, \alpha)$ satisfies Assumptions~\ref{assump-rate} and~\ref{assump-zero-entropy}. 

To check Assumption~\ref{assump-rate}, for $\alpha \in \mathbb{R}_{+*}$, we have:
\begin{align}
\frac{1}{n} \log \frac{1}{T( \mu^n, \alpha)(x^n)}  &= \frac{1}{n} \log \frac{\sum_{y^n \in \mc{X}^n} [\mu^n(y^n)]^\alpha}{[\mu^n(x^n)]^\alpha} \nonumber \\
& =\frac{1}{n} \left[\log \sum_{y^n \in \mc{X}^n} [\mu^n(y^n)]^\alpha +\log \frac{1}{[\mu^n(x^n)]^\alpha}\right]. \label{eq:bb-nobound}
\end{align}
Further, considering $\alpha \geq 1$,
\begin{equation}
- \alpha \log |\mc{X}| \leq (1- \alpha) \log |\mc{X}| \leq \frac{1}{n} \log  \sum_{y^n \in \mc{X}^n} [\mu^n(y^n)]^\alpha \leq 0,
\label{eq:bb-bound1}
\end{equation}
and for all $0<\alpha\leq 1$, 
\begin{align}
0 \leq \frac{1}{n} \log  \sum_{y^n \in \mc{X}^n} [\mu^n(y^n)]^\alpha \leq (1-\alpha)  \log |\mc{X}| \leq \log |\mc{X}|.
\label{eq:bb-bound2}
\end{align}
Thus, for all $\alpha \in \mathbb{R}_{+*}$, by taking the maximum of the upper bounds and the minimum of the lower bounds in~\eqref{eq:bb-bound1} and~\eqref{eq:bb-bound2}, and applying them to~\eqref{eq:bb-nobound}, we have
\begin{equation*}
- \alpha \log |\mc{X}| + \alpha \log \frac{1}{\overline{\varepsilon}} \leq  \frac{1}{n} \log\frac{1}{T(\mu^n, \alpha)(x^n)} \leq  \log |\mc{X}|  + \alpha \log \frac{1}{\underline{\varepsilon}}, 
\end{equation*}
where $\underline{\varepsilon}$ and $\overline{\varepsilon}$ are defined in~\eqref{eq:max-min-prob} and~\eqref{eq:min-max-prob}, respectively.
On the other hand, for all $\alpha \in \mathbb{R}_{-*}$, we have
\begin{align*}
\frac{1}{n} \log \frac{1}{T( \mu^n, \alpha)(x^n)}  & =\frac{1}{n} \left[\log \sum_{y^n \in \mc{X}^n} [\mu^n(y^n)]^\alpha +\log \frac{1}{[\mu^n(x^n)]^\alpha}\right]\nonumber \\
&\leq \frac{1}{n} \log \sum_{y^n \in \mc{X}^n} \underline{\varepsilon}^{n\alpha} \\
& = \log |\mc{X}| - \alpha \log \frac{1}{\underline{\varepsilon}}.
\end{align*}
Thus, for all $\alpha \in \mathbb{R}$, we have
\begin{equation}
 \frac{1}{n} \log\frac{1}{T(\mu^n, \alpha)(x^n)} \leq \log |\mc{X}| + |\alpha|\log \frac{1}{\underline{\varepsilon}} < \infty.
\label{eq:pppp1}
\end{equation}
%This implies that Assumption~\ref{assump-rate} is satisfied.
Next, we need to show that for all $\alpha \neq 0$,
\begin{equation*}
H(T (\mub, \alpha))>0.
\end{equation*}
Due to Lemma~\ref{lem:VV}, $H(T (\mub, \alpha))$ is differentiable with respect to $\alpha$ and its derivative is given by $\alpha V(T(\mub, \alpha)||\mub)$. On the other hand, for $\alpha > 0$, the derivative is strictly positive implying that $H(T(\mub, \alpha))$ is strictly decreasing with respect to $\alpha$, and hence it is non-zero for all $\alpha>0$.
This together with~\eqref{eq:pppp1} demonstrate that for all $\alpha \in \mathbb{R}$, $T(\mub, \alpha)$ satisfies Assumption~\ref{assump-rate}.

To show that Assumption~\ref{assump-zero-entropy} holds for the tilted source, we note that for any $\alpha > 0$, we have
\begin{align*}
\arg\min_{x \in \mc{X}} T(\mu^1, \alpha) (x) & = \arg\min_{x \in \mc{X}} \left\{ \frac{(\mu^1(x))^\alpha}{\sum_{y \in \mc{X}} (\mu^1(y))^\alpha}  \right\} \\
&= \arg\min_{x \in \mc{X}} (\mu^1(x))^\alpha \\
&= \arg\min_{x \in \mc{X}} \mu^1(x) ,
\end{align*}
which is unique by the fact that $\mub$ satisfies Assumption~\ref{assump-zero-entropy}. Similarly, we can argue that $\arg\max_{x \in \mc{X}} T(\mu^1, \alpha) (x)$  is also unique.
On the other hand, for $\alpha<0$, we have
\begin{align*}
\arg\min_{x \in \mc{X}} T(\mu^1, \alpha) (x) & = \arg\min_{x \in \mc{X}} \left\{ \frac{(\mu^1(x))^\alpha}{\sum_{y \in \mc{X}} (\mu^1(y))^\alpha}  \right\} \\
&= \arg\min_{x \in \mc{X}} (\mu^1(x))^\alpha\\
&= \arg\max_{x \in \mc{X}} \mu^1(x),
\end{align*}
which is again unique by the fact that $\mub$ satisfies Assumption~\ref{assump-zero-entropy}. Finally, $\arg\max_{x \in \mc{X}} T(\mu^1, \alpha) (x)$ is unique by the fact that $\arg\min_{x\in \mc{X}} \mu^1(x)$ is unique completing the proof.
\end{IEEEproof}

\section{Equivalent Order Class}
\label{sec:order_class}

In this section, we characterize the set of all string-sources  that induce the same optimal ordering of the strings (from the most likely to the least likely) on all strings of all lengths. 

\begin{definition}[order equivalent string-sources]
We say two string-sources $\mub, \rhob \in \mc{M}_{\mc{X}}$ are order equivalent and denote by $\mub \equiv \rhob$, if for all $n \in \mathbb{N}$, $G_{\mub}^n = G^n_{\rhob}$, i.e., the optimal ordering for $\mub$ is also the optimal ordering for $\rhob$. 
\end{definition}

{
Note that two string-sources may be order equivalent while they are not necessarily indistinguishable sources. By contrast, if for all $n \in \mathbb{N}$, $\mu^n = \rho^n,$ then the two are indistinguishable, and we write $\mub = \rhob$.
}

\begin{definition}[equivalent order class]
The equivalent order class of a string-source $\mub \in \mc{M}_{\mc{X}}$ is denoted by ${\mc{C}_{\mub}}$ and is given by
$
{\mc{C}_{\mub}} = \{ \rhob \in \mc{M}_{\mc{X}}:  \rhob \equiv \mub \}.
$
\label{def:equivalent_order}
\end{definition}

We also need another definition that generalizes the weakly typical set. 
\begin{definition}[tilted weakly typical set of order $\alpha$]
For any $\alpha \in \mathbb{R}$, denote by $\mc{A}^n_{\mub,\alpha,\epsilon}$ the tilted weakly typical set of order $\alpha$ with respect to $\mub$, which is the set of all $x^n$ such that for an $\epsilon \in \mathbb{R}_{+*}$:
\begin{equation}
e^{-H^n(T(\mub, \alpha)|| \mub) - n \epsilon }  < \mu^n(x^n) <  e^{-H^n(T(\mub, \alpha)|| \mub) + n \epsilon},
\label{eq:typical-set}
\end{equation}
where $H(\cdot|| \cdot)$ is the cross entropy defined in~\eqref{eq:HH}.
\label{def:typical-set}
\end{definition}
Note that in the case of $\alpha =1$, $\mc{A}^n_{\mub,1,\epsilon}$ is the familiar weakly typical set (see (3.6) in~\cite{cover-book}); and Definition~\ref{def:typical-set} generalizes the weakly typical set for $\alpha \neq 1$.
We shall establish some useful properties of the tilted weakly typical sets in the next section.

Our interest in finding the equivalent order class of a string-source $\mub$ lies in the fact that such string-sources are subject to the same optimal guessing procedure, which renders them equivalent in terms of many problems, such as brute-force password guessing, list decoding, sequential decoding, and one-to-one source coding.

\begin{theorem}
For any string-source $\mub \in \mc{M}_{\mc{X}}$, if $\rhob \in \mc{C}_\mub$, then there exists a unique $\alpha \in \mathbb{R}_{+*}$ such that 
$\rhob = T( \mub, \alpha)$. Consequently, 
the equivalent order class of $\mub$ is equal to the tilted family, i.e.,  ${\mc{C}_{\mub}} = \mc{T}^{+}_{\mub}.$
Further, the set of all string-sources whose optimal orderings are inverse for $\mub$ is equal to $\mc{T}^-_{\mub}$, i.e., $\mc{C}_{T ( \mub, -1)} = \mc{T}^-_{\mub}$.
\label{thm:order_class}
\end{theorem}

Theorem~\ref{thm:order_class} characterizes the set of all string-sources that would result in the same ordering of strings for any length. As discussed above, these string-sources are called order  equivalent as their optimal guesswork procedure is exactly the same. Theorem~\ref{thm:order_class} establishes that any such order equivalent string-source is necessarily the result of a tilt providing an operational meaning to the tilted family.

To prove the theorem, we state two intermediate lemmas and a definition.

\begin{lemma}
For any  string-source $\mub$, the equivalent order class of $\mub$ contains $\mc{T}^{+}_{\mub}$, i.e., $\mc{T}^{+}_{\mub} \subseteq {\mc{C}_{\mub}}$.
\label{lemma:17}
\end{lemma}
\begin{IEEEproof}
To prove the lemma, we need to show that for any $\alpha \in \mathbb{R}_{+*}$, we have $T (\mub, \alpha) \equiv \mub$, i.e., $ T (\mub, \alpha)  \in {\mc{C}_{\mub}}$. To this end, we show that  {$G^n_{\mub} = G^n_{T(\mub, \alpha)}$, i.e.,} $G^n_{\mub}$  is also the optimal ordering on $n$-strings for $T(\mub, \alpha)$.  Thus, we need to show that
\begin{equation}
T (\mu^n, {\alpha})(x^n) < T(\mu^n, {\alpha} )(y^n) \Rightarrow G^n_{\mub}(x^n) > G^n_{\mub}(y^n).
\end{equation}
{Considering~\eqref{eq:optimal_ordering}}, this establishes that $G^n_\mub$ is the optimal ordering for $T(\mu^n, \alpha)$, {and hence $G^n_\mub = G^n_{T(\mub, \alpha)}$.}
It suffices to show that for any $\alpha \in \mathbb{R}_{+*}$, we have
\begin{equation}
\mu^n(x^n) > \mu^n(y^n) \Leftrightarrow T( \mu^n, \alpha) (x^n) > T( \mu^n, \alpha) (y^n).
\end{equation}
We show here that 
\begin{align}
&\mu^n(x^n) > \mu^n(y^n)\nonumber\\ 
& \Leftrightarrow  [\mu^n(x^n)]^\alpha > [\mu^n(y^n)]^\alpha\label{eq:alpha-positive}\\
& \Leftrightarrow  \frac{[\mu^n(x^n)]^\alpha}{\sum_{z^n \in \mc{X}^n} [\mu^n(z^n)]^\alpha} > \frac{[\mu^n(y^n)]^\alpha}{\sum_{z^n \in \mc{X}^n} [\mu^n(z^n)]^\alpha} \nonumber\\
& \Leftrightarrow T( \mu^n, \alpha) (x^n) > T( \mu^n, \alpha) (y^n),\label{eq:stepp1}
\end{align}
where \eqref{eq:alpha-positive} holds because $\alpha >0$, and~\eqref{eq:stepp1} follows from the definition of the tilt in~\eqref{eq:tilt}.
\label{lem:contain}
\end{IEEEproof}

Next, we state a simple lemma about the weakly typical set (which is a special case of the tilted weakly typical sets).
\begin{lemma}
For any $\mub \in \mc{M}_{\mc{X}}$,
\begin{align}
\mathbb{P}_{\mub}\left\{X^n \in \mc{A}^n_{\mub,1,\epsilon} \right\} & \geq 1 - \frac{V^n(\mub)}{n^2 \epsilon^2},
\label{eq:CLT1}\\
|\mc{A}^n_{\mub,1,\epsilon} |   &> e^{H^n(\mub) - n \epsilon} \left(1 - \frac{V^n(\mub)}{n^2 \epsilon^2} \right),\label{eq:CLT1L}\\
|\mc{A}^n_{\mub,1,\epsilon} |  &<  e^{H^n(\mub) + n \epsilon}.\label{eq:CLT1U}
\end{align}
\label{lem:CLT}
\end{lemma}
\begin{IEEEproof}
\begin{align}
\mathbb{P}_{\mub}\left\{X^n \in \mc{A}^n_{\mub,1,\epsilon} \right\}  &= \mathbb{P}_{\mub}\left\{\left|\imath_\mub(X^n)  -H^n(\mub)\right| \leq n \epsilon  \right\}\nonumber\\
& \geq 1 - \frac{V^n(\mub)}{n^2 \epsilon^2}\label{eq:cheby},
\end{align}
where~\eqref{eq:cheby} follows from Chebyshev's inequality. Note that Chernoff bounds could be used to establish an exponential decay in~\eqref{eq:cheby} but the weaker bound above is sufficient for our purposes.
To derive the lower bound in~\eqref{eq:CLT1L} on the size of the set $\mc{A}_{\mub, 1, \epsilon}$, observe that
\begin{align*}
|\mc{A}_{\mub, 1, \epsilon}| &\geq  \frac{\mathbb{P}_\mub\{X^n \in \mc{A}_{\mub, 1, \epsilon}\}}{\max_{x^n \in \mc{A}_{\mub, 1, \epsilon}} \{\mu^n(x^n)\}}\\
&> \frac{1 - \frac{V^n(\mub)}{n^2 \epsilon^2} }{e^{-H^n(\mub) +n \epsilon}}\\
&= \left(1 - \frac{V^n(\mub)}{n^2 \epsilon^2} \right) e^{H^n(\mub) -n \epsilon}.
\end{align*}
The bound in~\eqref{eq:CLT1U} is established similarly.
\end{IEEEproof}

\begin{definition}[Shannon entropy contour]
Let $\mc{S}_h \subset \mc{M}_{\mc{X}}$ denote the Shannon entropy contour $h$, which is the set of all string-sources in $\mc{M}_{\mc{X}}$ whose entropy rate equals $h  \in (0,\log |\mc{X}|)$, i.e., 
\begin{equation}
\mc{S}_h := \{ \mub \in \mc{M}_{\mc{X}}: H(\mub) = h\}.
\end{equation}
\label{def:entropy_contour}
\end{definition}

Next, we provide the proof of the main result of this section on the equivalent order class of a string-source.
\begin{IEEEproof}[Proof of Theorem~\ref{thm:order_class}]
To prove the theorem, we take two steps. 
We first show that one element of the tilted family $\mc{T}_\mub^+$ is also contained in the intersection of the Shannon entropy contour and the equivalent order class, i.e., $
\mc{T}^+_{\mub} \cap {\mc{C}_{\mub}} \cap \mc{S}_h  \in \mc{M}_{\mc{X}}$ is not empty.
We also show that for any  string-source $\mub \in \mc{M}_{\mc{X}}$ and any $h \in (0,\log |\mc{X}|)$, if $\rhob_1, \rhob_2 \in \mc{M}_{\mc{X}}$ are in the intersection of $\mc{S}_h$ (see Definition~\ref{def:entropy_contour}) and ${\mc{C}_{\mub}}$ denoted by 
$
{\mc{C}_{\mub}} \cap \mc{S}_h,
$
then {$\rhob_1$ and $\rhob_2$ are indistinguishable sources, i.e., $\rhob_1 = \rhob_2$.  Putting these two steps together ensures that the intersection $
{\mc{C}_{\mub}} \cap \mc{S}_h
$
contains one and only one element, which is a member of the tilted family by Lemma~\ref{lemma:17}, and completes the proof. The second part of the  theorem is proved by invoking Lemma~\ref{lemma:reverse} and the first part of the theorem.}

We first show that ${\mc{C}_{\mub}} \cap \mc{S}_h$ is not empty and contains one element from $\mc{T}^+_{\mub}$. 
This is carried out by noting the continuity and monotonicity of $H(T(\mub, \alpha))$ in the parameter $\alpha$ proved in Lemma~\ref{lem:properties} and noting that $H(\mb{u}_{\mc{X}}) = \log |\mc{X}|$ and $\lim_{\alpha \to \infty} H^n(T( \mub, \alpha)) = 0$.
% Note that by Lemma~\ref{lem:Gamma_theta}, $\Gamma_\theta \subseteq {\mc{C}_{\mub}}$. Therefore, 

Next, we proceed the proof of uniqueness by contradiction.
Let $\rhob_1, \rhob_2 \in {\mc{C}_{\mub}} \cap \mc{S}_h$ be two string-sources that are contained in the intersection of $\mc{S}_h$ and ${\mc{C}_{\mub}}$. Note that by Definition~\ref{def:equivalent_order}, $\rhob_1, \rhob_2 \in \mc{M}_{\mc{X}}$.
Therefore, by definition as $\rhob_1, \rhob_2 \in \mc{S}_h$, we have 
\begin{equation*}
H(\rhob_1) = H(\rhob_2) =h,
\end{equation*}
where $h \in (0, \log |\mc{X}|)$ by Definition~\ref{def:entropy_contour}.
The claim is negated to assume that $\rhob_1$ and $\rhob_2$ are not equal, i.e., there exists $\delta \in \mathbb{R}_{+*}$ such that {for all $n \in \mathbb{N}$:
\begin{equation*}
D^n(\rhob_2||\rhob_1) = E_{\rhob_2}\left\{ \log \left( \frac{\rho_2^n(X^n)}{\rho_1^n(X^n)}\right)\right\} >2n \delta.
\end{equation*}
}
Then, we can decompose the expectation over the weakly typical set and the rest of the strings as follows:
\begin{align}
E_{\rhob_2} \left\{\log \left( \frac{\rho_2^n(X^n)}{\rho_1^n(X^n)}\right)\right\} & = \hspace{-.15in} \sum_{x^n \in \mc{A}^n_{\rhob_2, 1, \epsilon}} \hspace{-.15in}\rho_2^n(x^n) \log \left( \frac{\rho_2^n(x^n)}{\rho_1^n(x^n)}\right)\nonumber\\
&+ \hspace{-.15in}\sum_{x^n \not\in \mc{A}^n_{\rhob_2, 1, \epsilon}} \hspace{-.15in}\rho_2^n(x^n) \log \left( \frac{\rho_2^n(x^n)}{\rho_1^n(x^n)}\right).\label{eq:zero}
\end{align}
By Assumption~\ref{assump-rate} for the string-source $\rhob_1 \in \mc{M}_\mc{X}$, there exists $\underline{\varepsilon}$ such that
\begin{equation}
\max_{x^n \in \mc{X}^n} \log \frac{1}{\rho_1^n(x^n)}<n\log \frac{1}{\underline{\varepsilon}}. \label{fghy1}
\end{equation} 
On the other hand, by~\eqref{eq:CLT1} in Lemma~\ref{lem:CLT},
\begin{equation}
\sum_{x^n \not\in \mc{A}^n_{\rhob_2, 1, \epsilon}} \rho_2^n(x^n)  \leq  \frac{V^n(\rhob_2)}{n^2 \epsilon^2},\label{fghy2}
\end{equation}
Putting together~\eqref{fghy1} and~\eqref{fghy2}, we have
\begin{align}
\sum_{x^n \not\in \mc{A}^n_{\rhob_2, 1, \epsilon}}\rho_2^n(x^n) \log \left( \frac{\rho_2^n(x^n)}{\rho_1^n(x^n)}\right) \nonumber\\
\leq \sum_{x^n \not\in \mc{A}^n_{\rhob_2, 1, \epsilon}}\rho_2^n(x^n) \log \left( \frac{1}{\rho_1^n(x^n)}\right)\nonumber\\
 \leq \frac{V^n(\rhob_2)}{n \epsilon^2} \log \frac{1}{\underline{\varepsilon}} = \frac{V^1(\rhob_2)}{\epsilon^2} \log \frac{1}{\underline{\varepsilon}},
\end{align}
leading to the fact that the term in~\eqref{eq:zero} is $O_n(1)$.
Using the above in conjunction with the fact that $P_\mub\left\{X^n \in \mc{A}^n_{\rhob_2, 1, \epsilon} \right\} \geq 1 - \frac{V^n(\mub)}{n^2 \epsilon^2}$, 
we conclude that
for sufficiently large $n$: 
\begin{equation*}
E_{\rhob_2} \left\{ \left.\log \left( \frac{\rho_2^n(X^n)}{\rho_1^n(X^n)}\right)\right| X^n \in \mc{A}^n_{\rhob_2, 1, \epsilon} \right\}> n \delta.
\end{equation*}
Consequently, there exists $y^n \in \mc{A}^n_{\rhob_2, 1, \epsilon}$ such that 
\begin{equation}
\log \frac{1}{\rho_1^n(y^n)} > \log \frac{1}{\rho_2^n(y^n)} + n\delta.
\label{eq:rho-diff}
\end{equation}
Since  $y^n \in \mc{A}^n_{\rhob_2, 1, \epsilon}$, by definition, we have
\begin{align*}
nh - n\epsilon \leq \log \frac{1}{\rho_2^n(y^n)}  \leq nh + n\epsilon,
\end{align*}
which in conjunction with~\eqref{eq:rho-diff} leads to
\begin{equation*}
\log \frac{1}{\rho_1^n(y^n)} > \log \frac{1}{\rho_2^n(y^n)} + n\delta\geq nh-n\epsilon + n\delta.
\end{equation*}
By swapping $\rhob_1$ and $\rhob_2$ and noting that
\begin{equation*}
D(\rhob_2||\rhob_1) = 0 \Leftrightarrow D(\rhob_1||\rhob_2) = 0,
\end{equation*}
and repeating similar arguments, we could pick a different sequence $\hat{y}^n \in \mc{X}^n$ such that
\begin{equation*}
\log \frac{1}{\rho_2^n(\hat{y}^n)} >  nh-n\epsilon+ n\delta.
\end{equation*}
and 
\begin{align*}
\log \frac{1}{\rho_1^n(\hat{y}^n)}  \leq nh + n\epsilon.
\end{align*}
 Note that the above holds for any $\epsilon \in \mathbb{R}_{+*}$. By choosing $\epsilon$ to be small enough (smaller than $\delta/2$) we then ensure that for $n$ sufficiently large, $\rho_2^n(y^n)> \rho_2^n(\hat{y}^n)$ and hence $G^n_{\rhob_2}(y^n) < G^n_{\rhob_2}(\hat{y}^n)$. Similarly,
 we can show that for sufficiently large $n$, we have $\rho_1^n(y^n)< \rho_1^n(\hat{y}^n)$, which leads to $G_{\rhob_1}^n(y^n) > G_{\rhob_1}^n(\hat{y}^n)$. Therefore, $\rhob_1$ and $\rhob_2$ cannot be order equivalent, which is a contradiction. Hence, 
 it must be true that for any $\delta\in \mathbb{R}_{+*}$, {for sufficiently large $n$}:
 \begin{equation*}
E_{\rhob_2}\left\{ \log \left( \frac{\rho_2^n(X^n)}{\rho_1^n(X^n)}\right)\right\} <n \delta,
\end{equation*}
 which directly implies that
$ D(\rhob_1||\rhob_2) = D(\rhob_2||\rhob_1) = 0,$ and $\rhob_1 = \rhob_2$,
i.e., any two string-sources in the intersection of ${\mc{C}_{\mub}}$ and $\mc{S}_h$ must be equal, {which completes the proof of the uniqueness.}
\end{IEEEproof}

\section{Tilted Weakly Typical Set}
\label{sec:typical-set}

The main result in this section is a generalization of the asymptotic equipartition property on the weakly typical set concerning information and guesswork. We call this generalized notion the tilted weakly typical set and define it in Definition~\ref{def:typical-set} in the previous section. Although these results resemble the asymptotic equipartition property on the weakly typical set, they are concerned with the large deviations, i.e., atyptical behavior, rather than the typical behavior. We will consequently use these results in establishing the large deviation properties of guesswork and information in the next section.

We need two more definitions before we can provide the statement of the main result in this section.
\begin{definition}
For any $\mub \in \mc{M}_{\mc{X}}$, let $ \mc{B}^n_{\mub,\alpha ,\epsilon} \subset \mc{A}^n_{\mub,\alpha,\epsilon} $ be a subset such that $  \left|\mc{B}^n_{\mub,\alpha,\epsilon}\right| = \left\lfloor \frac{1}{2} \left|\mc{A}^n_{\mub,\alpha,\epsilon}\right|\right\rfloor$ and for any $x^n \in \mc{B}^n_{\mub,\alpha,\epsilon}$ and $y^n \in \mc{A}^n_{\mub,\alpha,\epsilon} \cap \overline{\mc{B}^n_{\mub,\alpha,\epsilon}}$ we have $T(\mu^n, \alpha)(x^n) < T(\mu^n, \alpha)(y^n)$.
\label{def-B}
\end{definition}

\begin{definition}
For any $\mub \in \mc{M}_{\mc{X}}$, let $ \mc{D}^n_{\mub,\alpha,\epsilon} \supseteq \mc{A}^n_{\mub,\alpha,\epsilon} $ be defined as follows:
\begin{equation}
\mc{D}^n_{\mub, \alpha, \epsilon} := \left\{ x^n \in \mc{X}^n : T(\mu^n, \alpha)(x^n) >  e^{- H^n(T(\mub, \alpha)) - n |\alpha |\epsilon}\right\}.
\label{eq:def-D}
\end{equation}
\label{def-D}
\end{definition}

\begin{definition}
For any $\mub \in \mc{M}_{\mc{X}}$, let $ \mc{E}^n_{\mub,\alpha,\epsilon} \supseteq \mc{A}^n_{\mub,\alpha,\epsilon} $ be defined as follows:
\begin{equation}
\mc{E}^n_{\mub, \alpha, \epsilon} := \left\{ x^n \in \mc{X}^n : T(\mu^n, \alpha) (x^n) <  e^{- H^n(T(\mub, \alpha) ) + n |\alpha| \epsilon}\right\}.
\label{eq:def-E}
\end{equation}
\label{def-D}
\end{definition}

\begin{theorem}
For any $\mub \in \mc{M}_{\mc{X}}$, any $\epsilon \in \mathbb{R}_{+*}$, and any $ \alpha \neq 0$, \vspace{0.07in} 

{\it (a)} For any $x^n \in \mc{A}^n_{\mub,\alpha,\epsilon}$, 
\begin{align}
\mu^n(x^n) &>e^{-H^n(T( \mub, \alpha)|| \mub)-n \epsilon},\label{eq:LDP-i-lower}\\
\mu^n(x^n) &< e^{-H^n(T( \mub, \alpha)|| \mub)+n \epsilon};
\label{eq:LDP-i-upper}
\end{align}

{\it (b)} For any $n \in \mathbb{N}$,
\begin{align}
| \mc{A}^n_{\mub,\alpha,\epsilon}|&>\left(1 - \frac{V^n(T(\mub, \alpha)||\mub)}{n^2 \epsilon^2} \right)  e^{H^n(T( \mub, \alpha)) - |\alpha|n \epsilon} ,\label{eq:LDP-size-lower}\\
| \mc{A}^n_{\mub,\alpha,\epsilon}| &< e^{H^n(T( \mub, \alpha)) + |\alpha|n \epsilon};\label{eq:LDP-size-upper}
\end{align}

{\it (c)} For any $n \in \mathbb{N}$, and $\alpha \in (-\infty, 0) \cup [1,\infty)$,
\begin{align}
\mathbb{P}_{\mub}\left\{X^n \in \mc{D}^n_{\mub,\alpha,\epsilon} \right\} &\geq \mathbb{P}_{\mub}\left\{X^n \in \mc{A}^n_{\mub,\alpha,\epsilon} \right\}\nonumber\\
 >  \left(1 - \frac{V^n(T(\mub, \alpha)||\mub)}{n^2 \epsilon^2} \right) &e^{-D^n(T(\mub, \alpha)||\mub) - |1-\alpha| n \epsilon } ,\label{eq:LDP-P-lower}\\
\mathbb{P}_{\mub}\left\{X^n \in \mc{A}^n_{\mub,\alpha,\epsilon} \right\} &\leq \mathbb{P}_{\mub}\left\{X^n \in \mc{D}^n_{\mub,\alpha,\epsilon} \right\}\nonumber\\
&\leq e^{-D^n(T(\mub, \alpha)||\mub) + |1-\alpha| n \epsilon} ,
\label{eq:LDP-P-upper}\\
 \mathbb{P}_{\mub}\left\{X^n  \in \mc{E}^n_{\mub,\alpha,\epsilon} \right\}&\geq  1- e^{-D^n(T(\mub, \alpha)||\mub) + |1-\alpha| n \epsilon} ;
\end{align}

{\it (d)} For any $n \in \mathbb{N}$, and $\alpha \in (0,1]$,
\begin{align}
\mathbb{P}_{\mub}\left\{X^n \in \mc{E}^n_{\mub,\alpha,\epsilon} \right\} &\geq \mathbb{P}_{\mub}\left\{X^n \in \mc{A}^n_{\mub,\alpha,\epsilon} \right\}\nonumber\\
 >  \left(1 - \frac{V^n(T(\mub, \alpha)||\mub)}{n^2 \epsilon^2} \right)&e^{-D^n(T(\mub, \alpha)||\mub) - |1-\alpha| n \epsilon } ,\label{eq:LDP-P2-lower}\\
\mathbb{P}_{\mub}\left\{X^n \in \mc{A}^n_{\mub,\alpha,\epsilon} \right\} &\leq \mathbb{P}_{\mub}\left\{X^n \in \mc{E}^n_{\mub,\alpha,\epsilon} \right\}\nonumber\\
&\leq e^{-D^n(T(\mub, \alpha)||\mub) + |1-\alpha| n \epsilon},
\label{eq:LDP-P2-upper}\\
\mathbb{P}_{\mub}\left\{X^n  \in \mc{D}^n_{\mub,\alpha,\epsilon} \right\}&\geq  1-e^{-D^n(T(\mub, \alpha)||\mub) + |1-\alpha| n \epsilon} ;
\end{align}

{\it (e)} For any $n \in \mathbb{N}$ and any $\alpha>0$:
\begin{align}
x^n \in \mc{B}^n_{\mub,\alpha,\epsilon} \quad &\Rightarrow\nonumber\\
 G_\mub^n(x^n) &> \frac{1}{2} \left(1 - \frac{V^n(T(\mub, \alpha)||\mub)}{n^2 \epsilon^2} \right) e^{H^n(T(\mub, \alpha)) -\alpha n \epsilon},\label{eq:thm2-g-1}\\
x^n \in \mc{D}^n_{\mub,\alpha,\epsilon} \quad &\Rightarrow \quad G_\mub^n(x^n) \leq  e^{H^n(T(\mub, \alpha)) + \alpha n\epsilon},\label{eq:thm2-g-2}\\
x^n \in \mc{D}^n_{\mub,\alpha,\epsilon}\quad&\Leftarrow \nonumber\\
G_\mub^n(x^n) &\leq  \left(1 - \frac{V^n(T(\mub, \alpha)||\mub)}{n^2 \epsilon^2} \right) e^{H^n(T(\mub, \alpha)) -\alpha n \epsilon},\label{eq:thm2-g-3}\\
x^n \in \mc{E}^n_{\mub,\alpha,\epsilon}\quad&\Leftarrow \quad G_\mub^n(x^n) > e^{H^n(T(\mub, \alpha)) +\alpha n \epsilon}.\label{eq:thm2-g-4}
\end{align}

{\it (f)} For any $n \in \mathbb{N}$ and any $\alpha<0$:
\begin{align}
x^n \in \mc{B}^n_{\mub,\alpha,\epsilon} \quad &\Rightarrow \nonumber\\
 R_\mub^n(x^n) & > \frac{1}{2} \left(1 - \frac{V^n(T(\mub, \alpha)||\mub)}{n^2 \epsilon^2} \right) e^{H^n(T(\mub, \alpha)) -|\alpha| n \epsilon},\label{eq:thm2-r-1}\\
x^n \in \mc{D}^n_{\mub,\alpha,\epsilon} \quad &\Rightarrow \quad R_\mub^n(x^n) \leq  e^{H^n(T(\mub, \alpha)) + |\alpha| n\epsilon},\label{eq:thm2-r-2}\\
x^n \in \mc{D}^n_{\mub,\alpha,\epsilon}\quad&\Leftarrow \nonumber\\
 R_\mub^n(x^n) &\leq \left(1 - \frac{V^n(T(\mub, \alpha)||\mub)}{n^2 \epsilon^2} \right) e^{H^n(T(\mub, \alpha)) -|\alpha| n \epsilon},\label{eq:thm2-r-3}\\
x^n \in \mc{E}^n_{\mub,\alpha,\epsilon}\quad&\Leftarrow \quad R_\mub^n(x^n) > e^{H^n(T(\mub, \alpha)) +|\alpha| n \epsilon}.\label{eq:thm2-r-4}
\end{align}
\label{thm:LDP-information}
\end{theorem}

Recall that Shannon entropy, cross entropy, and relative entropy are measured in nats throughout this paper.
We first provide a few lemmas that specialize to the typical set, which we use to prove the theorem.
\begin{lemma}
For any $\mub \in \mc{M}_{\mc{X}}$, let $x^n \in \mc{B}^n_{\mub,1,\epsilon}$ (see Definition~\ref{def-B}). Then, 
\begin{align}
G_\mub^n(x^n) &> \frac{1}{2} \left(1 - \frac{V^n(\mub)}{n^2 \epsilon^2} \right) e^{H^n(\mub) -n \epsilon}.
\end{align}
\label{lem:typical-1}
\end{lemma}
\begin{IEEEproof}
Note that by definition for all $x^n \in \mc{B}^n_{\mub, 1, \epsilon}$, we have $G_\mub^n(x^n) > |\overline{ \mc{B}^n_{\mub, 1, \epsilon}}|\geq \frac{1}{2}  |\mc{A}^n_{\mub,1,\epsilon}|$. The proof is completed noticing the bound in~\eqref{eq:CLT1L}.
\end{IEEEproof}

\begin{lemma}
For any $\mub \in \mc{M}_{\mc{X}}$, let $x^n \in \mc{D}^n_{\mub, 1, \epsilon}$ (see Definition~\ref{def-D}). Then, 
\begin{align}
G_\mub^n(x^n) &\leq  e^{H^n(\mub) + n\epsilon}.
\end{align}
\label{lem:typical-2}
\end{lemma}
\begin{IEEEproof}
We have
\begin{align}
G_\mub^n(x^n) &\leq |\mc{D}^n_{\mub, 1, \epsilon}|\label{eq:139}\\
 & \leq \frac{1}{\min_{x^n \in \mc{D}^n_{\mub, 1, \epsilon}}\left\{\mu^n(x^n)\right\}} \label{eq:140}\\
&= e^{H^n( \mub) + n \epsilon},\label{eq:141}
\end{align}
where~\eqref{eq:139} follows from the definition in~\eqref{eq:def-D},~\eqref{eq:140} follows from the fact that the probability of the set $\mc{D}^n_{\mub, 1, \epsilon}$ is upper bounded by $1$, and~\eqref{eq:141} again follows from the definition in~\eqref{eq:def-D}.
\end{IEEEproof}

\begin{lemma}
Suppose the following holds:
\begin{equation}
G_\mub^n(x^n) \leq \left(1 - \frac{V^n(\mub)}{n^2 \epsilon^2} \right) e^{H^n(\mub) -n \epsilon}.
\label{eq:lem-typical-3}
\end{equation}
Then, $x^n \in \mc{D}^n_{\mub, 1, \epsilon}$.
\label{lem:typical-3}
\end{lemma}
\begin{IEEEproof}
We proceed the proof with contradiction. Suppose that $x^n \not\in \mc{D}^n_{\mub, 1, \epsilon}.$ Hence, $\mu^n(x^n) \leq e^{-n H(\mub) -n\epsilon}$. Thus, $\mu^n(x^n)<\mu^n(y^n)$ for all  $y^n \in \mc{A}^n_{\mub, 1, \epsilon}$. Consequently, 
\begin{equation*}
G_\mub^n(x^n) > \left| \mc{A}^n_{\mub, 1, \epsilon} \right| >   \left(1 - \frac{V^n(\mub)}{n^2 \epsilon^2} \right) e^{H^n(\mub) -n \epsilon},
\end{equation*}
which contradicts the supposition of the lemma in~\eqref{eq:lem-typical-3}.
\end{IEEEproof}

\begin{lemma}
Suppose the following holds:
\begin{equation}
G_\mub^n(x^n) > e^{H^n(\mub) +n \epsilon}.
\label{eq:lem-typical-4}
\end{equation}
Then, $x^n \in \mc{E}^n_{\mub, 1, \epsilon}$.
\label{lem:typical-4}
\end{lemma}
\begin{IEEEproof}
We proceed the proof with contradiction. Suppose that $x^n \not\in \mc{E}^n_{\mub, 1, \epsilon}.$ Hence, $\mu^n(x^n) \geq e^{-n H(\mub) + n\epsilon}$. Thus, $\mu^n(x^n)>\mu^n(y^n)$ for all  $y^n \in \mc{A}^n_{\mub, 1, \epsilon}$. Consequently, 
\begin{equation*}
G_\mub^n(x^n) < G_\mub^n(y^n)\leq  e^{H^n(\mub) +n \epsilon},
\end{equation*}
which contradicts the supposition of the lemma in~\eqref{eq:lem-typical-4}.
\end{IEEEproof}

Next, we provide a lemma of equivalence between tilted weakly typical sets and weakly typical sets of a tilted source, which we will rely on in proving the main theorem of this section.
\begin{lemma}
Let $\mub \in \mc{M}_\mc{X}$. Then, for all $\alpha \neq 0$, the following equivalences hold: 
\begin{align}
\mc{A}^n_{\mub, \alpha, \epsilon} &= \mc{A}^n_{T(\mub,\alpha),1, |\alpha| \epsilon},\label{eq:equi1}\\
\mc{B}^n_{\mub, \alpha, \epsilon} &= \mc{B}^n_{T(\mub,\alpha),1, |\alpha| \epsilon},\label{eq:equi2}\\
\mc{D}^n_{\mub, \alpha, \epsilon} &= \mc{D}^n_{T(\mub,\alpha),1, |\alpha| \epsilon}, \label{eq:equi3}\\
\mc{E}^n_{\mub, \alpha, \epsilon} &= \mc{E}^n_{T(\mub,\alpha),1, |\alpha| \epsilon}. \label{eq:equi4}
\end{align}
\label{lem:typical-equivalence}
\end{lemma}
\begin{IEEEproof}
We have
\begin{align}
\mc{A}^n_{T(\mub,\alpha),1, |\alpha| \epsilon} &= \left\{ x^n \in \mc{X}^n:  \left|\log \frac{1}{T(\mu^n, \alpha)(x^n)} - H^n(T(T(\mub,\alpha),1)||T(\mub,\alpha))\right| <|\alpha|n \epsilon\right\} \label{eq:try1}\\ 
& = \left\{ x^n \in \mc{X}^n:  \left| \frac{1}{\alpha}\log \frac{1}{T(\mu^n, \alpha)(x^n)} - \frac{1}{\alpha}H^n(T(\mub,\alpha))\right| <n \epsilon\right\}  \label{eq:try2}\\
& = \left\{ x^n \in \mc{X}^n:  \left| \log \frac{1}{\mu^n(x^n)} - H^n(T(\mub,\alpha)||\mub)\right| < n \epsilon\right\} \label{eq:try3} \\
& = \mc{A}^n_{\mub, \alpha, \epsilon},\nonumber
\end{align}
where~\eqref{eq:try1} follows from Definition~\ref{def:typical-set},~\eqref{eq:try2} follows by noting that $T(\mub, 1) = \mub$ and $H^n(\mub||\mub) = H^n(\mub)$ and the fact that $\alpha \neq 0$, and~\eqref{eq:try3} follows from Lemma~\ref{lem:tilt-mean}.

Next, note that~\eqref{eq:equi2} follows directly from~\eqref{eq:equi1} and Definition~\ref{def-B}.

To establish~\eqref{eq:equi3} note that:
\begin{align}
\mc{D}^n_{T(\mub,\alpha),1, |\alpha| \epsilon} &= \left\{ x^n \in \mc{X}^n:  \log \frac{1}{T(\mu^n, \alpha)(x^n)} - H^n(T(T(\mub,\alpha),1)||T(\mub,\alpha)) <|\alpha| n \epsilon\right\} \label{eq:try1-2}\\ 
& = \left\{ x^n \in \mc{X}^n: \log \frac{1}{T(\mu^n, \alpha)(x^n)} - H^n(T(\mub,\alpha)) <|\alpha|n \epsilon\right\}  \label{eq:try2-2}\\
& = \mc{D}^n_{\mub, \alpha, \epsilon},\nonumber
\end{align}
where~\eqref{eq:try1-2} follows from Definition~\ref{def:typical-set},~\eqref{eq:try2-2} follows by noting that $T(\mub, 1) = \mub$ and $H^n(\mub||\mub) = H^n(\mub)$ and the fact that $\alpha \neq 0$.
Finally, the proof for~\eqref{eq:equi4} is very similar to that of~\eqref{eq:equi3} with all the inequalities reversed, and is omitted for brevity.
\end{IEEEproof}
Next, we proceed to the proof of Theorem~\ref{thm:LDP-information}.
\begin{IEEEproof}[Proof of Theorem~\ref{thm:LDP-information}]
Part (a) is simply a reiteration of Definition~\ref{def:typical-set}, which is presented here for the completeness of the characterization.

To establish Part (b), 
notice that by Lemma~\ref{lem:typical-equivalence}, for all $\alpha \neq 0$, we have $\mc{A}^n_{T(\mub,\alpha),1, |\alpha| \epsilon}  = \mc{A}^n_{\mub, \alpha, \epsilon}$.
Part (b) is established by applying the bounds in~\eqref{eq:CLT1L} and~\eqref{eq:CLT1U} in Lemma~\ref{lem:CLT} to $\mc{A}^n_{T(\mub,\alpha),1, |\alpha| \epsilon}$, and noting that $\alpha^2 V(T(\mub, \alpha)||\mub) = V(T(\mub, \alpha))$ from Lemma~\ref{lem:mismatched-varentropy}.

In Parts (c) and (d), first notice that by definition $\mc{A}^n_{\mub, \alpha, \epsilon} \subseteq \mc{D}^n_{\mub, \alpha, \epsilon}$ and $\mc{A}^n_{\mub, \alpha, \epsilon} \subseteq \mc{E}^n_{\mub, \alpha, \epsilon}$ and hence 
$\mathbb{P}_{\mub}\left\{X^n \in \mc{A}^n_{\mub,\alpha,\epsilon} \right\} \leq \mathbb{P}_{\mub}\left\{X^n \in \mc{D}^n_{\mub,\alpha,\epsilon} \right\}$ and $\mathbb{P}_{\mub}\left\{X^n \in \mc{A}^n_{\mub,\alpha,\epsilon} \right\} \leq \mathbb{P}_{\mub}\left\{X^n \in \mc{E}^n_{\mub,\alpha,\epsilon} \right\}$. Hence, we only need to prove the right hand side inequalities in the claims.

To prove~\eqref{eq:LDP-P-lower} in Part (c), consider~\eqref{eq:CLT1} in Lemma~\ref{lem:CLT} applied to $\mc{A}^n_{T(\mub,\alpha),1, |\alpha| \epsilon}$, in conjunction with~\eqref{eq:equi1} in Lemma~\ref{lem:typical-equivalence}: 
\begin{align}
\mathbb{P}_{T(\mub, \alpha)}\left\{X^n \in \mc{A}^n_{\mub,\alpha,\epsilon} \right\}  &= \mathbb{P}_{T(\mub, \alpha)}\left\{\mc{A}^n_{T(\mub,\alpha),1, |\alpha| \epsilon}\right\} \nonumber\\
&\geq  1 - \frac{V^n(T(\mub, \alpha)||\mub)}{n^2 \epsilon^2}.
\label{eq:prf1}
\end{align}
On the other hand, we have 
\begin{align}
\mathbb{P}_{T(\mub, \alpha)}\left\{X^n \in \mc{A}^n_{\mub,\alpha,\epsilon} \right\} &= \sum_{x^n \in \mc{A}^n_{\mub,\alpha,\epsilon}} T(\mu^n, \alpha) (x^n)\nonumber\\
& = \frac{\sum_{x^n \in \mc{A}^n_{\mub,\alpha,\epsilon}} [ \mu^n(x^n)]^\alpha}{\sum_{x^n \in \mc{X}^n} [ \mu^n(x^n)]^\alpha}\label{eq:sss1}\\
& = \frac{\sum_{x^n \in \mc{A}^n_{\mub,\alpha,\epsilon}}\mu^n(x^n) [ \mu^n(x^n)]^{\alpha-1}}{e^{(1-\alpha) H_\alpha^n(\mub)}}\label{eq:sss2}\\
& < e^{(1-\alpha)\left(H^n(T(\mub, \alpha)|| \mub) -H_\alpha^n(\mub)\right)+ |\alpha-1|n \epsilon}\sum_{x^n \in \mc{A}^n_{\mub,\alpha,\epsilon}}\mu^n(x^n) \label{eq:sss3}\\
& = e^{D^n(T(\mub, \alpha)|| \mub)+ |\alpha-1|n \epsilon}\mathbb{P}_{\mub}\left\{X^n \in \mc{A}^n_{\mub,\alpha,\epsilon} \right\},
\label{eq:prf2}
\end{align}
where~\eqref{eq:sss1} follow from the definition of the tilt in~\eqref{eq:tilt},~\eqref{eq:sss2} follows from the definition of the R\'enyi entropy in~\eqref{eq:renyi_entropy},~\eqref{eq:sss3} follows from~\eqref{eq:LDP-i-lower} or~\eqref{eq:LDP-i-upper} applied to $[\mu^n(x^n)]^{\alpha-1}$ depending on the sign of $(\alpha-1)$, and~\eqref{eq:prf2} follows from~\eqref{eq:renyi-eq1} in Lemma~\ref{lem:renyi-eq}.  
Putting~\eqref{eq:prf1} and~\eqref{eq:prf2} together we arrive at the right hand side inequality in~\eqref{eq:LDP-P-lower}.  Note that all the steps are applicable for $\alpha \neq 0$, and hence are also applicable to the case where $\alpha \in (0,1]$. This also establishes the right hand side inequality in~\eqref{eq:LDP-P2-upper} for Part (d).

In Part (c), the upper bound in~\eqref{eq:LDP-P-upper} would be established as follows. We first note:
\begin{equation}
\mathbb{P}_{T(\mub, \alpha)}\left\{X^n \in \mc{D}^n_{\mub,\alpha,\epsilon} \right\}  = \mathbb{P}_{T(\mub, \alpha)}\left\{\mc{D}^n_{T(\mub,\alpha),1, |\alpha| \epsilon}\right\} \leq  1 .
\label{eq:prf1-2-5}
\end{equation}
We also note that~\eqref{eq:def-D} in conjunction with Lemma~\ref{lem:tilt-mean} imply that for $\alpha>0$:
\begin{equation}
\mc{D}^n_{\mub, \alpha, \epsilon} = \left\{ x^n \in \mc{X}^n : \mu^n(x^n) >  e^{- H^n(T(\mub, \alpha)||\mub) - n \epsilon}\right\}.
\label{eq:def-D1}
\end{equation}
and for $\alpha<0$:
\begin{equation}
\mc{D}^n_{\mub, \alpha, \epsilon} = \left\{ x^n \in \mc{X}^n :\mu^n(x^n) <  e^{- H^n(T(\mub, \alpha)||\mub) + n \epsilon}\right\}.
\label{eq:def-D2}
\end{equation}
Thus,
\begin{align}
\mathbb{P}_{T(\mub, \alpha)}\left\{X^n \in \mc{D}^n_{\mub,\alpha,\epsilon} \right\} & = \frac{\sum_{x^n \in \mc{D}^n_{\mub,\alpha,\epsilon}}\mu^n(x^n) [ \mu^n(x^n)]^{\alpha-1}}{e^{(1-\alpha) H_\alpha^n(\mub)}}\label{eq:sss2-2}\\
& > e^{(1-\alpha)\left(H^n(T(\mub, \alpha)|| \mub) -H_\alpha^n(\mub)\right)- |\alpha-1|n \epsilon}\sum_{x^n \in \mc{D}^n_{\mub,\alpha,\epsilon}}\mu^n(x^n) \label{eq:sss3-2}\\
& = e^{D^n(T(\mub, \alpha)|| \mub)- |\alpha-1|n \epsilon}\mathbb{P}_{\mub}\left\{X^n \in \mc{D}^n_{\mub,\alpha,\epsilon} \right\}\label{eq:prf2-2},
\end{align}
where~\eqref{eq:sss3-2} follows from~\eqref{eq:def-D1} (for $\alpha>1$) and~\eqref{eq:def-D2} (for $\alpha<0$) applied to $[\mu^n(x^n)]^{\alpha-1}$, and~\eqref{eq:prf2-2} follows from~\eqref{eq:renyi-eq1} in Lemma~\ref{lem:renyi-eq}.  
Putting~\eqref{eq:prf1-2-5} and~\eqref{eq:prf2-2} together we arrive at the desired upper bound in~\eqref{eq:LDP-P-upper}. 
The upper bound in~\eqref{eq:LDP-P2-upper} is achieved by repeating~\eqref{eq:sss2-2}--\eqref{eq:prf2-2} for $\mc{E}^n_{\mub,\alpha,\epsilon}$ in the case of $\alpha \in (0,1]$ and applying~\eqref{eq:def-E}.

To establish Parts (e) and (f), we first apply Lemmas~\ref{lem:typical-1}--\ref{lem:typical-4}, to the sets $\mc{B}^n_{T(\mub,\alpha),1, |\alpha| \epsilon}$, $\mc{D}^n_{T(\mub,\alpha),1, |\alpha| \epsilon}$, and $\mc{E}^n_{T(\mub,\alpha),1, |\alpha| \epsilon}$ and notice the equivalences in Lemma~\ref{lem:typical-equivalence} to get:
\begin{align*}
x^n \in \mc{B}^n_{\mub,\alpha,\epsilon} \quad &\Rightarrow \quad G_{T(\mub, \alpha)}^n(x^n) \geq \frac{1}{2} \left(1 - \frac{V^n(T(\mub, \alpha))}{n^2 \alpha^2\epsilon^2} \right) e^{H^n(T(\mub, \alpha)) -|\alpha| n \epsilon},\\
x^n \in \mc{D}^n_{\mub,\alpha,\epsilon} \quad &\Rightarrow \quad G_{T(\mub, \alpha)}^n(x^n) <   e^{H^n(T(\mub, \alpha)) + |\alpha| n\epsilon},\\
x^n \in \mc{D}^n_{\mub,\alpha,\epsilon}\quad&\Leftarrow \quad G_{T(\mub, \alpha)}^n(x^n) <  \left(1 - \frac{V^n(T(\mub, \alpha))}{n^2 \alpha^2\epsilon^2} \right) e^{H^n(T(\mub, \alpha)) +|\alpha| n \epsilon},\\
x^n \in \mc{E}^n_{\mub,\alpha,\epsilon}\quad&\Leftarrow \quad G_{T(\mub, \alpha)}^n(x^n) > e^{H^n(T(\mub, \alpha)) + |\alpha| n \epsilon}.
\end{align*}
Parts (e) and (f) then follow by invoking Theorem~\ref{thm:order_class} to notice that for all $x^n \in \mc{X}^n$, $ G_{T(\mub, \alpha)}^n(x^n) = G_\mub^n(x^n)$ for $\alpha>0$ and $ G_{T(\mub, \alpha)}^n(x^n) = R_\mub^n(x^n)$ for $\alpha<0$, and noting that $\alpha^2 V(T(\mub, \alpha)||\mub) = V(T(\mub, \alpha))$ from Lemma~\ref{lem:mismatched-varentropy}.
\end{IEEEproof}

Theorem~\ref{thm:LDP-information} (a)-(d), for $\alpha = 1$, reduces to the properties of the weakly typical set for finite-alphabet memoryless sources (see~\cite[Theorem 3.1.2]{cover-book}).
Part (e) for $\alpha = 1$ provides a characterization of guesswork for weakly typical strings.
For values of $\alpha \neq 1,$ the theorem further provides an asymptotic equipartition property for atypical strings generated by memoryless sources.
It is well known that several information theoretic quantities, such as the average length of the optimal source coding, and the channel capacity, could be analyzed through the typical or jointly typical sequences, and the entropy rate of the process appears as the fundamental performance limit. On the other hand, several other important notions, such as the error exponents are governed by the rare events comprising of atypical behavior. 
For different values of $\alpha$, Theorem~\ref{thm:LDP-information} describes the large deviation performance of information and the logarithm of (forward) guesswork, which generalizes the notion of the typical set. In particular, it also recovers the large deviations performance of the optimal length function for prefix-free source coding as well. We remark that Theorem~\ref{thm:LDP-information} could also be used to recover known results that are concerned with rare events in a more intuitive way.
Finally, we expect Theorem~\ref{thm:LDP-information} to be generalizable to a winder class of stationary ergodic sources in light of the generalized version of the asymptotic equipartition property proved by Bobkov and Madiman~\cite{mokshay-AEP2, mokshay-AEP}.

Note that the tighter non-asymptotic bounds on the moments of guesswork derived in~\cite{sason-guesswork} could not be used to fully characterize the distribution of guesswork as in Theorem~\ref{thm:LDP-information}. In particular, for the unlikely half of the sequences, the characterization involves reverse guesswork which is first studied in this paper.

\section{Large Deviations}
\label{sec:LDP}
Thus far, we provided an operational meaning of the tilted family of the string-source in Theorem~\ref{thm:order_class}, and generalized the notion of the weakly typical set to the tilted weakly typical sets that characterize all strings in Theorem~\ref{thm:LDP-information}.
Before we use these tools to provide results on large deviations, we need one more definition.

\begin{definition}[Large Deviation Principle]
A sequence of random variables $\{Y_n:n\in\mathbb{N}\}$ taking values in $\mathbb{R}$ satisfies the Large Deviation Principle with rate function $J:\mathbb{R}\to[0,\infty]$ if $J$ is lower semi-continuous and has compact level-sets, and for all Borel sets $B$
\begin{align*}
-\inf_{t\in B^\circ} J(t)
        \leq \liminf_{n\to\infty} \frac 1n \log \mathbb{P}\{Y_n\in B\}
        \leq \limsup_{n\to\infty} \frac 1n \log \mathbb{P}\{Y_n\in B\}
        \leq -\inf_{t\in \bar{B}} J(t),
\end{align*}
where $B^\circ$ is the interior of $B$ and $\bar{B}$ is its closure.
\end{definition}

If the $Y_n$ take values in a compact subset $K$ of $\mathbb{R}$, then to establish that $\{Y_n: n \in \mathbb{N}\}$ satisfies the LDP it is sufficient \cite[Theorem 4.1.11 and Lemma 1.2.18]{dembo-book} to prove that the following exists for all $t\in K$
\begin{align*}
\lim_{\epsilon\downarrow0} \liminf_{n\to\infty} \frac 1n \log \mathbb{P}\left\{Y_n\in (t-\epsilon, t+\epsilon)\right\}
        =\lim_{\epsilon\downarrow0} \limsup_{n\to\infty} \frac 1n \log \mathbb{P}\left\{Y_n\in (t-\epsilon, t+\epsilon)\right\}
        = - J(t),
\end{align*}
where $\downarrow$ and $\uparrow$ denote approaching from above and from below, respectively.
Note that all of the processes of interest to us take values in compact sets, since for all $x^n \in \mc{X}^n$,
\begin{align*}
\frac{1}{n} g_\mub^n(x^n) \in [0, \log |\mc{X}|],\quad\quad\quad
\frac{1}{n} r_\mub^n(x^n)  \in [0, \log |\mc{X}|], \quad\quad\quad
\frac{1}{n} \imath_\mub^n(x^n)  \in  \left[\log \frac{1}{\overline{\varepsilon}} , \log \frac{1}{\underline{\varepsilon}}\right].
\end{align*}
Thus, for our purposes it is sufficient to prove that
\begin{align}
\lim_{\epsilon \downarrow  0} \lim_{n \to \infty}\frac{1}{n}\log \left(\frac{1}{\mathbb{P} \left\{|Y_n -t |<\epsilon \right\}}\right) = J(t).
\end{align}

Christiansen and Duffy proved that the logarithm of (forward) guesswork, $\left\{\frac{1}{n}g_\mub^n(X^n) : n \in \mathbb{N} \right\},$ defined in Definition~\ref{def:exponent} satisfies a LDP and characterized the rate function as the solution to an optimization problem~\cite{duffy-LDP}, which is the Legendre-Fenchel transform of the scaled cumulant generating function.
In the rest of this section, we characterize the large deviations behavior of reverse guesswork, guesswork, and information, respectively, and provide their rate functions in terms of information theoretic quantities.

\subsection{Reverse guesswork}
First, we use the generalization of the typical set and its properties, established in the previous section, to prove that the logarithm of reverse guesswork satisfies a LDP. We emphasize that the method used by Christiansen and Duffy~\cite{duffy-LDP} would not lead to proving a LDP for the logarithm of reverse guesswork  because the resulting rate function is non-convex as will be proved later in this section.

\begin{theorem}
For any $\mub \in \mc{M}_{\mc{X}}$, the sequence $\left\{\frac{1}{n} r_\mub^n(x^n): n\in \mathbb{N} \right\}$ satisfies a LDP, and the corresponding rate function is implicitly given by
\begin{equation}
J^r_\mub(t) = D(T(\mub, \alpha_\mub^r(t))|| \mub),
\label{eq:}
\end{equation}
where
\begin{equation}
\alpha_\mub^r(t) = \arg_{\alpha \in \mathbb{R}_{-*}} \left\{ H(T(\mub, \alpha)) = t\right\}.
\label{eq:a-tr}
\end{equation}
\label{thm:rate-r}
\end{theorem}

\begin{IEEEproof}
Consider $\alpha_\mub^r(t)$ defined in~\eqref{eq:a-tr} implying that $H(T(\mub, \alpha_\mub^r(t))) = t$. Hence, for sufficiently large $n$,
\begin{align}
\mathbb{P}_\mub \left\{\left|\frac{1}{n} \log R_\mub^n(X^n) - t\right| < \epsilon\right\} &=\mathbb{P}_\mub \left\{\left|\frac{1}{n} \log R_\mub^n(X^n) - H(T(\mub, \alpha_\mub^r(t))) \right| < \epsilon\right\} \nonumber\\
&= \mathbb{P}_\mub \left\{\left| \log R_\mub^n(X^n) - H^n(T(\mub, \alpha_\mub^r(t)))\right| < n\epsilon\right\} \nonumber\\
&\leq \mathbb{P}_\mub \left\{ X^n \in \mc{D}^n_{\mub, \alpha_\mub^r(t), (2\epsilon) / \alpha_\mub^r(t)}\right\} \label{eq:bbb3} \\
& \leq e^{-D^n(T(\mub, \alpha_\mub^r(t))||\mub) + 2\frac{|1-\alpha_\mub^r(t)|}{\alpha_\mub^r(t)}n \epsilon  }  \label{eq:bbb4},
\end{align}
where~\eqref{eq:bbb3} holds because~\eqref{eq:thm2-r-3} ensures that for sufficiently large $n$ we have
\begin{equation*}
\left\{x^n: \left| \log R_\mub^n(x^n) - H^n(T(\mub, \alpha_\mub^r(t)))\right| < n\epsilon \right\} \subseteq \mc{D}^n_{\mub, \alpha_\mub^r(t), (2\epsilon) / \alpha_\mub^r(t)},
\end{equation*}
and hence 
\begin{equation*}
P_\mub\left\{x^n: \left| \log R_\mub^n(x^n) - H^n(T(\mub, \alpha_\mub^r(t)))\right| < n\epsilon \right\} \leq P_\mub\left\{ X^n \in \mc{D}^n_{\mub, \alpha_\mub^r(t), (2\epsilon) / \alpha_\mub^r(t)}\right\},
\end{equation*}
and~\eqref{eq:bbb4} is a direct application of the upper bound in~\eqref{eq:LDP-P2-upper} to $\mc{D}^n_{\mub, \alpha_\mub^r(t), 2\epsilon / \alpha_\mub^r(t)}.$
Consequently,
\begin{equation*}
\frac{1}{n}\log \left(\frac{1}{\mathbb{P}_\mub \left\{\left|\frac{1}{n} \log R_\mub^n(X^n) - t\right| < \epsilon\right\}}\right) \geq \frac{1}{n}D^n(T(\mub, \alpha_\mub^r(t))||\mub) - 2\frac{|1-\alpha_\mub^r(t)|}{\alpha_\mub^r(t)}\epsilon.
\end{equation*}
Taking limits as $n\to \infty$ and letting $\epsilon \downarrow 0$ would lead to
\begin{equation}
\lim_{\epsilon \downarrow 0} \lim_{n\to \infty}\frac{1}{n}\log \left(\frac{1}{\mathbb{P}_\mub \left\{\left|\frac{1}{n} \log R_\mub^n(X^n) - t\right| < \epsilon\right\}}\right) \geq D(T(\mub, \alpha_\mub^r(t))||\mub) .
\label{eq:asdf1}
\end{equation}
Next, we will show that the left hand side of~\eqref{eq:asdf1} is also upper bounded by $D(T(\mub, \alpha_\mub^r(t))||\mub)$. We have
\begin{align}
\mathbb{P}_\mub \left\{\left|\frac{1}{n} \log R_\mub^n(X^n) - t\right| < \epsilon\right\} 
&=\mathbb{P}_\mub \left\{\left| \log R_\mub^n(X^n) - nH(T(\mub, \alpha_\mub^r(t))) \right| < n\epsilon\right\} \nonumber\\
&\geq \mathbb{P}_\mub \left\{ X^n \in \mc{B}^n_{\mub, {\alpha_\mub^r(t)}, \epsilon /(2{\alpha_\mub^r(t)})}\right\} \label{eq:bbbb3} \\
& \geq e^{-D^n(T(\mub, \alpha_\mub^r(t))||\mub) - \frac{|1-{\alpha_\mub^r(t)}|}{2{\alpha_\mub^r(t)}}n \epsilon  } \label{eq:bbbb4},
\end{align}
where~\eqref{eq:bbbb3} holds because~\eqref{eq:thm2-r-1} and~\eqref{eq:thm2-r-2} ensure that 
\begin{equation*}
\mc{B}^n_{\mub, \alpha_\mub^r(t), \epsilon / (2\alpha_\mub^r(t))} \subseteq \left\{x^n: \left| \log R_\mub^n(x^n) - H^n(T(\mub, \alpha_\mub^r(t)))\right| < n\epsilon \right\},
\end{equation*}
and~\eqref{eq:bbbb4} is due to the lower bound in~\eqref{eq:LDP-P-lower}. Hence, following steps similar to the previous case, %and noting that $$\lim_{n \to \infty} \log (1+o_n(1)) = 0,$$
 we have
\begin{equation*}
\lim_{\epsilon \downarrow 0} \lim_{n\to \infty}\frac{1}{n}\log \left(\frac{1}{\mathbb{P}_\mub \left\{\left|\frac{1}{n} \log R_\mub^n(X^n) - t\right| < \epsilon\right\}}\right) \leq D(T(\mub, \alpha_\mub^r(t))||\mub),
\end{equation*}
which in conjunction with~\eqref{eq:asdf1} completes the proof.
\end{IEEEproof}
The parametric representation of the rate function allows us to readily deduce some of the properties of the rate function.

\begin{theorem}
The LDP rate function $J^r_\mub(t)$ associated with the logarithm of the reverse guesswork is a concave function of $t$, with its first and second derivatives given by
\begin{align}
&\frac{d}{dt} J_\mub^r(t) = \frac{1- \alpha_\mub^r(t)}{\alpha_\mub^r(t)},\label{eq:prop-r-1}\\
&\frac{d^2}{dt^2}J_\mub^r(t) = \frac{1}{\alpha_\mub^r(t) V(T(\mub, \alpha_\mub^r(t) ))}\label{eq:prop-r-2}.
\end{align}
It also satisfies the following properties:
\begin{align}
&\lim_{t \downarrow 0} \frac{d}{dt} J_\mub^r(t) = -1,\label{eq:prop-r-3}\\
& \lim_{t \uparrow \log|\mc{X}|} \frac{d}{dt} J_\mub^r(t) = -\infty,\label{eq:prop-r-4}\\
&\lim_{t \downarrow 0} \frac{d^2}{dt^2} J_\mub^r(t) = 0,\label{eq:prop-r-5}\\
& \lim_{t \uparrow \log|\mc{X}|} \frac{d^2}{dt^2} J_\mub^r(t) = -\infty.\label{eq:prop-r-6}
\end{align}
\label{thm:property-r}
\end{theorem}
\begin{IEEEproof}
To establish~\eqref{eq:prop-r-1}, note that since $J^r_\mub(\cdot)$ could be implicitly defined as given in Theorem~\ref{thm:rate-r}, and hence its derivative is given by the chain rule
\begin{align}
\frac{d}{dt} J_\mub^r(t) &= \frac{d \alpha_\mub^r(t)}{dt} \left.\frac{d}{d\alpha} D(T(\mub, \alpha)|| \mub)\right|_{\alpha = \alpha_\mub^r(t)}\label{eq:chain_rule}\\
&  = \left. \frac{\frac{d}{d \alpha} D(T(\mub, \alpha)|| \mub) }{\frac{d}{d \alpha} H(T(\mub, \alpha))} \right|_{\alpha = \alpha_\mub^r(t)}\nonumber\\
& = \frac{\alpha_\mub^r(t)-1}{\alpha_\mub^r(t)},
\label{eq:step}
\end{align}
where~\eqref{eq:chain_rule} follows from the chain rule, and~\eqref{eq:step} follows from the implicit function theorem implying that
\begin{equation}
\frac{d \alpha_\mub^r(t)}{dt}  = \left. \frac{1 }{\frac{d}{d \alpha} H(T(\mub, \alpha))} \right|_{\alpha = \alpha_\mub^r(t)}.
\label{eq:implicit_function}
\end{equation}
Invoking Lemmas~\ref{lem:H_decrease} and~\ref{lem:D_derivative} to calculate the numerator and denominator in~\eqref{eq:step} would lead to the desired result in~\eqref{eq:prop-r-1}.

To establish~\eqref{eq:prop-r-2}, note that 
\begin{align}
\frac{d^2}{dt^2} J_\mub^r(t) &= \frac{d}{dt} \left\{\frac{d}{dt} J_\mub^r(t)\right\}\nonumber\\
& =  \frac{d}{dt} \left\{ \frac{\alpha_\mub^r(t)-1}{\alpha_\mub^r(t)} \right\}\label{eq:199}\\
& = \frac{d\alpha_\mub^r(t)}{d t}   \left.  \frac{d}{d\alpha} \left\{ \frac{\alpha-1}{\alpha}\right\} \right|_{\alpha = \alpha_\mub^r(t)}  \label{eq:200} \\
& = \left.  \frac{ \frac{d}{d\alpha}  \frac{\alpha-1}{\alpha}}{\frac{d}{d \alpha} H(T(\mub, \alpha))} \right|_{\alpha = \alpha_\mub^r(t)}\nonumber\\
& =  \frac{1}{\alpha_\mub^r(t) V(T(\mub, \alpha_\mub^r(t) ))},
\label{eq:step2}
\end{align}
where~\eqref{eq:199} follows from~\eqref{eq:prop-r-1},~\eqref{eq:200} follows from the application of the implicit function theorem in~\eqref{eq:implicit_function}, and~\eqref{eq:step2} follows from Lemma~\ref{lem:H_decrease}, leading to the desired result in~\eqref{eq:prop-r-2}.

Equations \eqref{eq:prop-r-3} and \eqref{eq:prop-r-5} are derived by noting that the limit as $t \downarrow 0$, would be equivalent to the limit as $\alpha \to -\infty$ in the parametric form.
Similarly, equations \eqref{eq:prop-r-4} and \eqref{eq:prop-r-6} are derived by noting that the limit as $t \uparrow \log |\mc{X}|$, would be equivalent to the limit as $\alpha \uparrow 0$ in the parametric form.
Finally note that the concavity of $J^r_\mub(t)$ is evident as $\frac{d^2}{dt^2}J_\mub^r(t) >0$ from~\eqref{eq:prop-r-2} for all $\alpha <0$.
\end{IEEEproof}

\subsection{Guesswork}
The same proof methodology used for the logarithm of reverse guesswork leads us to establish a LDP for the logarithm of (forward) guesswork. Although this result is already proved by Christiansen and Duffy~\cite{duffy-LDP}, we further provide the implicit characterization of the rate function of the logarithm of guesswork  for all string-sources in $\mc{M}_{\mc{X}}$ using information theoretic quantities. %, which was already known by Christiansen and Duffy~\cite{duffy-LDP}. We further relate the rate function to the tilted family of the string-source.

\begin{theorem}
For any $\mub \in \mc{M}_{\mc{X}}$, the sequence $\left\{\frac{1}{n} g_\mub^n(X^n): n\in \mathbb{N} \right\}$ satisfies the large deviations principle with rate function $J^g_\mub$. The rate function is implicitly given by
\begin{equation}
J^g_\mub(t) = D(T(\mub, \alpha_\mub^g(t))|| \mub),
\end{equation}
where
\begin{equation}
\alpha_\mub^g(t) := \arg_{\alpha \in \mathbb{R}_{+*}} \left\{ H(T(\mub, \alpha)) = t\right\}.
\label{eq:a-t}
\end{equation}
Further, 
$J^g_\mub(0) = \lim_{\alpha \to \infty} D(T(\mub, \alpha)|| \mub)$
and 
$J^g_\mub(\log |\mc{X}|) = D(\mb{u}_\mc{X}|| \mub)$,
where $\mb{u}_{\mc{X}}$ is the uniform string-source on the alphabet $\mc{X}$ (see~\eqref{eq:def-uniform}).
\label{thm:rate-g}
\end{theorem}
\begin{IEEEproof}
The steps of the proof are very similar to the steps of the proof of Theorem~\ref{thm:rate-r} except that in this case $\alpha >0$, and we will be invoking the bounds on $G_\mub^n$ in~\eqref{eq:thm2-g-1}--\eqref{eq:thm2-g-3}. We will also need to separate the $\alpha \in (0,1)$ case and use the bounds on $\mc{E}^n_{\mub, \alpha, \epsilon}$ instead of $\mc{D}^n_{\mub, \alpha, \epsilon}$. 
\end{IEEEproof}

Next, we use Theorem~\ref{thm:rate-g} to establish some properties of the rate function for the logarithm of guesswork.

\begin{theorem}
The rate function $J^g_\mub(\cdot)$ associated with the logarithm of guesswork is a convex function of its argument, and particularly its first and second derivatives are given by
\begin{align}
&\frac{d}{dt} J_\mub^g(t) = \frac{1- \alpha_\mub^g(t)}{\alpha_\mub^g(t)},\label{eq:prop-g-1}\\
&\frac{d^2}{dt^2}J_\mub^g(t) = \frac{1}{\alpha_\mub^g(t) V(T(\mub, \alpha_\mub^g(t) ))}.\label{eq:prop-g-2}
\end{align}
Further, the following properties are satisfied:
\begin{align}
&\lim_{t \downarrow 0} \frac{d}{dt} J_\mub^g(t) = -1,\\
& \lim_{t \uparrow \log|\mc{X}|} \frac{d}{dt} J_\mub^g(t) = +\infty,\\
&\lim_{t \downarrow 0} \frac{d^2}{dt^2} J_\mub^g(t) = 0,\\
& \lim_{t \uparrow \log|\mc{X}|} \frac{d^2}{dt^2} J_\mub^g(t) = +\infty.
\end{align}
\label{thm:property-g}
\end{theorem}
\begin{IEEEproof}
The proof steps are similar to that of the proof of Theorem~\ref{thm:property-r}. We only point out the differences here. The limit as $t \uparrow \log |\mc{X}|$, would be equivalent to the limit as $\alpha \downarrow 0$ in the parametric form. The covexity of $J^g_\mub(t)$ is proved by noting that $\frac{d^2}{dt^2}J_\mub^g(t) >0$ from~\eqref{eq:prop-g-2} for all $\alpha >0$.
\end{IEEEproof}

\subsection{Information}
Next, we state a result on the large deviations of information. This result could be viewed as a generalization of the Shannon-McMillan-Breiman Theorem~\cite{Shannon-McMillan, Barron-AEP} and the characterization of the exponential rate of decay for the probability of atypical strings (see~\cite{mokshay-AEP} for a survey on the recent results on that front).

\begin{theorem}
For any $\mub \in \mc{M}_{\mc{X}}$, the sequence $\left\{\frac{1}{n} \imath_\mub^n(X^n): n\in \mathbb{N} \right\}$ satisfies a LDP, and the rate function is implicitly given by
\begin{equation}
J^\imath_\mub(t) = D(T(\mub, \alpha^\imath_\mub(t))|| \mub),
\end{equation}
where
\begin{equation}
\alpha^\imath_\mub(t) = \arg_{\alpha \in \mathbb{R}} \left\{ H(T(\mub, \alpha)||\mub) = t\right\}.
\label{eq:a-i}
\end{equation}
\end{theorem}

\begin{IEEEproof}
First notice that for any $\alpha \neq 0$ we have:
\begin{align}
\left\{x^n: \left|\frac{1}{n} \imath_\mub^n(x^n) - H(T(\mub, \alpha)||\mub) \right| < \epsilon\right\} &= \left\{x^n: e^{-H^n(T( \mub, \alpha)|| \mub)-n \epsilon} <\mu^n(x^n) < e^{-H^n(T( \mub, \alpha)|| \mub)+n \epsilon}\right\}\nonumber\\
& = \left\{ x^n: x^n \in \mc{A}^n_{\mub, \alpha, \epsilon} \right\}.
\label{eq:set-equal2}
\end{align}
Now, consider $\alpha_\mub^\imath(t)$ defined in~\eqref{eq:a-i} implying that $H(T(\mub, \alpha_\mub^\imath(t)||\mub) = t$.
\begin{align}
\mathbb{P}_\mub \left\{\left|\frac{1}{n} \imath_\mub^n(X^n) - t\right| < \epsilon\right\} &=\mathbb{P}_\mub \left\{\left|\frac{1}{n} \imath_\mub^n(X^n) - H(T(\mub, \alpha_\mub^\imath(t))||\mub) \right| < \epsilon\right\} \nonumber\\
&= \mathbb{P}_\mub \left\{ X^n \in \mc{A}^n_{\mub, \alpha_\mub^\imath(t), \epsilon}\right\} \label{eq:bbbc3} \\
& < e^{-D^n(T(\mub, \alpha_\mub^\imath(t))||\mub) + |1-\alpha_\mub^\imath(t)| n \epsilon  }  \label{eq:bbbc4},
\end{align}
where~\eqref{eq:bbbc3} follows from~\eqref{eq:set-equal2}, 
and~\eqref{eq:bbbc4} is a direct application of the upper bound in~\eqref{eq:LDP-P-upper}.
Consequently,
\begin{equation*}
\frac{1}{n}\log \left(\frac{1}{\mathbb{P}_\mub \left\{\left|\frac{1}{n} \imath_\mub^n(X^n) - t\right| < \epsilon\right\}}\right) \geq \frac{1}{n}D^n(T(\mub,\alpha_\mub^\imath(t))||\mub) - |1-\alpha_\mub^\imath(t)|\epsilon.
\end{equation*}
Taking limits as $n\to \infty$ and letting $\epsilon \downarrow 0$ would lead to
\begin{equation}
\lim_{\epsilon \downarrow 0} \lim_{n\to \infty}\frac{1}{n}\log \left(\frac{1}{\mathbb{P}_\mub \left\{\left|\frac{1}{n}  \imath_\mub^n(X^n) - t\right| < \epsilon\right\}}\right) \geq D(T(\mub, \alpha_\mub^\imath(t))||\mub).
\label{eq:asdf1-2}
\end{equation}
The rest of the proof entails showing that the above limit is upper bounded by $D(T(\mub, \alpha_\mub^\imath(t))||\mub).$ 
We have
\begin{align}
\mathbb{P}_\mub \left\{\left|\frac{1}{n} \imath_\mub^n(X^n) - t\right| < \epsilon\right\} 
&= \mathbb{P}_\mub \left\{ X^n \in \mc{A}^n_{\mub, {\alpha_\mub^\imath(t)}, \epsilon }\right\} \nonumber \\
& > \left(1 - \frac{V^n(T(\mub, \alpha)||\mub)}{n^2 \epsilon^2} \right) e^{-D^n(T(\mub, \alpha_\mub^\imath(t))||\mub) - |1-{\alpha_\mub^\imath}(t)| n \epsilon  }  \label{eq:bbbb4-2},
\end{align}
where~\eqref{eq:bbbb4-2} is a direct application of the upper bound in~\eqref{eq:LDP-P-lower}. 
Hence, following steps similar to the previous case, %and noting that $$\lim_{n \to \infty} \log (1+o_n(1)) = 0,$$
 we have
\begin{equation*}
\lim_{\epsilon \downarrow 0} \lim_{n\to \infty}\frac{1}{n}\log \left(\frac{1}{\mathbb{P}_\mub \left\{\left|\frac{1}{n}  \imath_\mub^n(X^n) - t\right| < \epsilon\right\}}\right) \leq D(T(\mub, \alpha_\mub^\imath(t))||\mub).
\end{equation*}
which in conjunction with~\eqref{eq:asdf1-2} completes the proof.
%whose steps are similar to that of Theorem~\ref{thm:rate-g} and the above, and are omitted for brevity.
\end{IEEEproof}

The above theorem on large deviations of information also lets us establish some properties of information.
\begin{definition}[min(max)-cross entropy]
For any $\mub \in \mc{M}_{\mc{X}}$, the min-cross entropy is defined as
\begin{equation}
t_\mub^- : = \lim_{\alpha \to +\infty} H(T(\mub, \alpha)||\mub),
\end{equation}
and the max-cross entropy  is defined as
\begin{equation}
t_\mub^+ : = \lim_{\alpha \to -\infty} H(T(\mub, \alpha)||\mub).
\end{equation}
\label{def-t}
\end{definition}
Roughly speaking, the likelihood of the most likely string scales like $e^{-t_\mub^- n}$. In words, the likelihood of the most likely string scales exponentially with $n$ and the exponent of the scaling is the negative min-cross entropy. Similarly, the likelihood of the least likely string scales as $e^{-t_\mub^+ n}$.
By definition, for any $x^n \in \mc{X}^n$, up to a multiplicative $(1+ o_n(1))$ term, we have
\begin{equation}
t_\mub^- \leq \frac{1}{n}\log \frac{1}{\mu^n(x^n)} \leq t_\mub^+.
\end{equation}
Thus, the normalized negative log-likelihood is bounded between the min-cross entropy and the max-cross entropy.
Equipped with this definition, we will characterize the properties of the rate function for information in the next theorem.

\begin{theorem}
The information rate function $J^\imath_\mub$ is a convex function, and its first two derivatives are given by
\begin{align}
&\frac{d}{dt} J_\mub^\imath(t) = 1- \alpha_\mub^\imath(t),\label{eq:prop-i-1}\\
&\frac{d^2}{dt^2}J_\mub^\imath(t) = \frac{\alpha_\mub^\imath(t)^2}{ V(T(\mub, \alpha_\mub^\imath(t) ))}.\label{eq:prop-i-2}
\end{align}
Further, it satisfies the following properties:
\begin{align}
&\lim_{t \downarrow t_\mub^-} \frac{d}{dt} J_\mub^\imath(t) = -\infty,\\
&\lim_{t \to \log |\mc{X}|} \frac{d}{dt} J_\mub^\imath(t) = 1,\\
&\lim_{t \uparrow t_\mub^+} \frac{d}{dt} J_\mub^\imath(t) = + \infty,\\
& \lim_{t \downarrow t_\mub^-} \frac{d^2}{dt^2} J_\mub^\imath(t) = +\infty,\\
&\lim_{t \to \log |\mc{X}|} \frac{d^2}{dt^2} J_\mub^\imath(t) =\frac{1}{V (\mb{u}_\mc{X} ||\mub)},\\
& \lim_{t \uparrow t_\mub^+} \frac{d^2}{dt^2} J_\mub^\imath(t) = +\infty,
\end{align}
where $t_\mub^-$ and $t_\mub^+$ are the min-cross entropy and the max-cross entropy, respectively, as in Definition~\ref{def-t}.
\label{thm:property-i}
\end{theorem}
\begin{IEEEproof}
To establish~\eqref{eq:prop-i-1}, note that  $J^\imath_\mub$ could be implicitly defined as given in Theorem~\ref{thm:rate-g}, and thus its derivative would be given by
\begin{equation}
\frac{d}{dt} J_\mub^\imath(t) = \left. \frac{\frac{d}{d \alpha} D(T(\mub, \alpha)|| \mub) }{\frac{d}{d \alpha} H(T(\mub, \alpha) || \mub)} \right|_{\alpha = \alpha_\mub^\imath(t)},
\label{eq:step3}
\end{equation}
invoking the implicit function theorem similar to the proof of Theorem~\ref{thm:property-r}.
Invoking Lemmas~\ref{lem:H_monotone} and~\ref{lem:D_derivative} to calculate the numerator and denominator in~\eqref{eq:step3} would lead to~\eqref{eq:prop-i-1}.

To establish~\eqref{eq:prop-i-2}, note that applying the implicit function theorem we get
\begin{equation}
\frac{d^2}{dt^2} J_\mub^\imath(t) = \frac{d}{dt} \frac{d}{dt} J_\mub^\imath(t) = \left.  \frac{ \frac{d}{d\alpha}  (1-\alpha)}{\frac{d}{d \alpha} H(T(\mub, \alpha)||\mub)} \right|_{\alpha = \alpha_\mub^\imath(t)}.
\label{eq:step4}
\end{equation}
The desired result is obtained by applying Lemma~\ref{lem:H_monotone} to the denominator of~\eqref{eq:step4}.

The rest of the properties are established by noting the definition of $t_\mub^-$ and $t_\mub^+$ in Definition~\ref{def-t} and considering the limits as $\alpha \to+ \infty$ and $\alpha \to -\infty$, respectively.
\end{IEEEproof}

\section{Approximation of the Probability Mass Function of Guesswork}
\label{sec:approximation}

Thus far, we provided some results on the asymptotic scaling of guesswork. In this section, we would like to provide more practical results that would be applicable for small sequences in practice.
Inspired from the characterization of guesswork using the tilt, we provide an approximation formula that is accurate even when the length $n$ is small. Our justification mostly relies on Theorem~\ref{thm:LDP-information} where we employ additional insights to compensate the impact of the finite sequence length.
The main contribution of this section is the following set of approximation formulas. 
\begin{approximation}
For any $\mub \in \mc{M}_{\mc{X}}$, the size of the tilted weakly typical set of order $\alpha \neq 0$ can be approximated by:
\begin{equation}
|\mc{A}^n_{\mub, \alpha, \epsilon}| \approx \frac{1 - e^{-2\alpha n \epsilon}}{\sqrt{2\pi V^n(T(\mub, \alpha))}} e^{H^n(T(\mub, \alpha)) +\alpha n \epsilon}
\end{equation}
\end{approximation}

\begin{approximation}
For any $\mub \in \mc{M}_{\mc{X}}$, and for any $x^n$ such that $\mu^n(x^n) = e^{-H^n(T(\mub, \alpha)||\mub)}$ for some $\alpha \in \mathbb{R}_{+*}$, we can approximately express its guesswork as
\begin{equation}
G^n_\mub(x^n) \approx \frac{e^{H^n(T(\mub, {\alpha} ))}}{\sqrt{\frac{\pi}{2}V^n(T(\mub, \alpha))} + \sqrt{\frac{\pi}{2} V^n(T(\mub, \alpha)) + 4}}.
\label{eq:approx1}
\end{equation}
Furthermore, for any $x^n$ such that $\mu^n(x^n) = e^{-H^n(T(\mub, \alpha)||\mub)}$ for some $\alpha \in \mathbb{R}_{-*}$, the reverse guesswork is approximately given by
\begin{equation}
R^n_\mub(x^n) \approx \frac{e^{H^n(T(\mub, {\alpha} ))}}{\sqrt{\frac{\pi}{2}V^n(T(\mub, \alpha))} + \sqrt{\frac{\pi}{2} V^n(T(\mub, \alpha)) + 4}}.
\label{eq:approx2}
\end{equation}
Alternatively, guesswork can be approximated by
\begin{align}
G^n_\mub(x^n) &= |\mc{X}|^n - R^n_\mub(x^n) \nonumber\\
&\approx |\mc{X}|^n - \frac{e^{H^n(T(\mub, {\alpha} ))}}{\sqrt{\frac{\pi}{2}V^n(T(\mub, \alpha))} + \sqrt{\frac{\pi}{2} V^n(T(\mub, \alpha)) + 4}} .
\label{eq:approx22}
\end{align}
\label{approx1}
Finally, for any $x^n$ such that $\mu^n(x^n) = e^{-H^n(\mathbf{u}_{\mc{X}} ||\mub)}$, the guesswork would be approximated as 
\begin{equation}
G^n_\mub(x^n) \approx  \frac{1}{2} |\mc{X}|^n.
\label{eq:approx23}
\end{equation}
\end{approximation}
\begin{IEEEproof}[Justification]
We have
\begin{align}
G^n_\mub(x^n) &\approx \int_{0}^{\infty} \frac{1}{\sqrt{2\pi V^n(T(\mub, \alpha))}}e^{H^n(T(\mub, {\alpha} )) - t - \frac{1}{2 V^n(T(\mub, \alpha))} t^2} dt \label{eq:laplace}\\
& = \frac{1}{2}  \exp\left({\frac{1}{2}V^n(T(\mub, \alpha))}\right) \text{erfc}\left(\sqrt{\frac{1}{2} V^n(T(\mub, \alpha))}\right) \label{eq:erfc}\exp\left(H^n(T(\mub, {\alpha} ))\right)\\
& \approx \frac{e^{H^n(T(\mub, {\alpha} ))}}{\sqrt{\frac{\pi}{2}V^n(T(\mub, \alpha))} + \sqrt{\frac{\pi}{2} V^n(T(\mub, \alpha)) + 4}},\label{eq:frank}
\end{align}
where~\eqref{eq:laplace} follows because of the Laplace's method,~\eqref{eq:erfc} follows the definition of erfc$(\cdot)$, and~\eqref{eq:frank} follows by applying the lower bound in~\cite{frank-note}.
\end{IEEEproof}

%\subsection{Numerical examples}
In the rest of this section, we provide numerical examples to show the effectiveness of the approximation formulas in Approximation~\ref{approx1}. We will also plot their asymptotic properties, such as the large deviations performance of the logarithm of guesswork using the formulas derived in the previous sections.
To this end, we formally define the four string-sources that will be used in the numerical experiments. The subscript in the following definitions denotes the alphabet size.

\begin{figure*}
\centering
\includegraphics[width=0.495\textwidth]{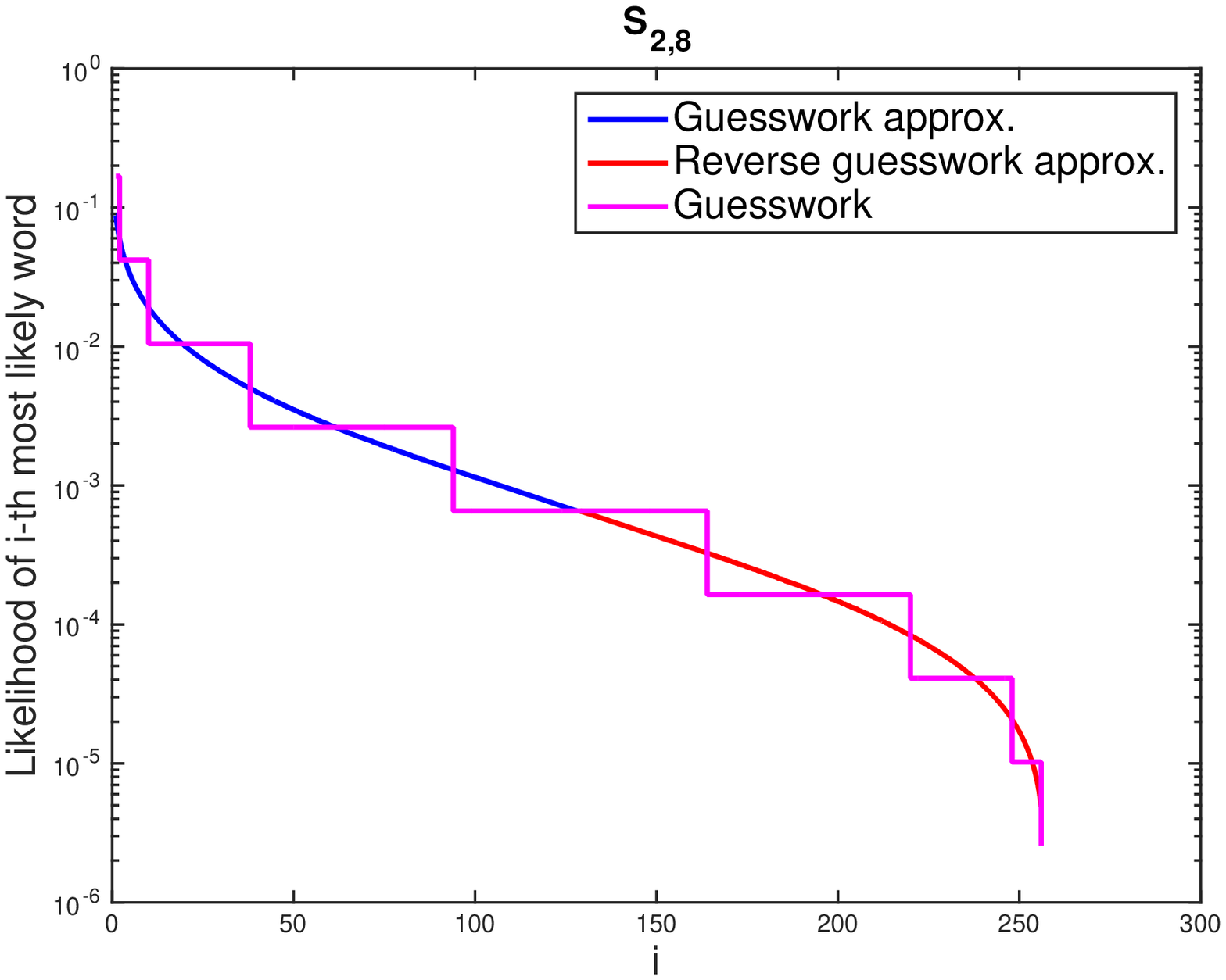}
\includegraphics[width=0.495\textwidth]{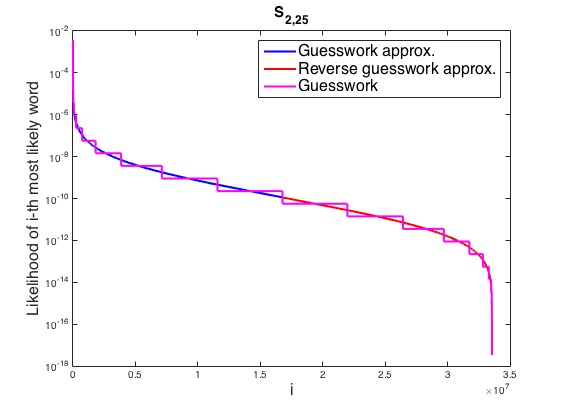}\\
\includegraphics[width=0.495\textwidth]{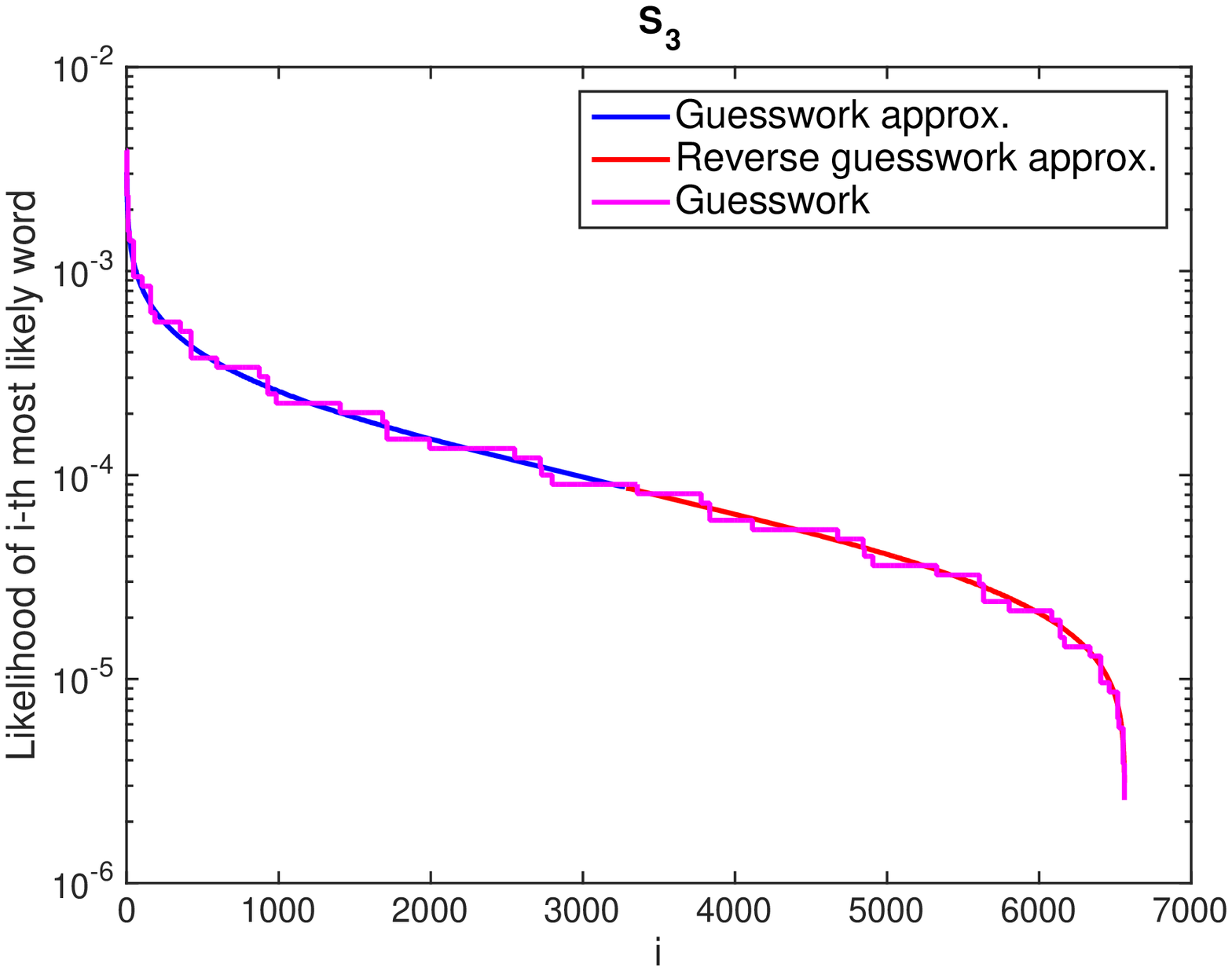}
\includegraphics[width=0.495\textwidth]{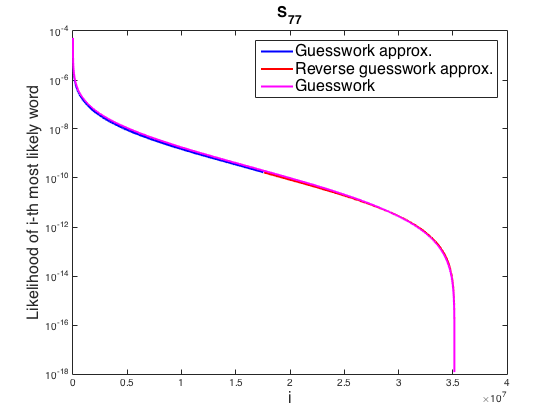}\\
\caption{The approximation on the distribution of the guesswork for string-sources $S_2$, $S_3$, and $S_{77}$. Note that $S_{2,8}$ denotes strings of length $n=8$ and $S_{2,25}$ denotes strings of length $n=25$. The string length for $S_3$ and $S_{77}$ are chosen to be $n=8$ and $n=4$ respectively. The pink curve represents the probability mass function of guesswork on the $|\mc{X}|^n$ strings; the blue curve is the approximation derived from the forward path in~\eqref{eq:approx1}; and the red curve is the approximation derived from the reverse path in~\eqref{eq:approx22}.}
\label{fig:guesswork-approx1}
\end{figure*}

The first example is a binary string-source.
\begin{definition}[string-source $S_{8}$]
Let $S_2$ refer to a Bernoulli source on alphabet on a binary alphabet with parameter vector (categorical distribution) $\theta = (0.2, 0.8)$.
We will use the notation $S_{2,n}$ to denote a source with output of length $n$.
\end{definition}

The next string-source is the ternary memoryless source used  as the working example throughout the paper.
\begin{definition}[string-source $S_3$]
Let $S_3$ refer to a memoryless string-source alphabet $\mc{X} = \{a,b,c \}$ of size $|\mc{X}| =3$ with parameter vector (categorical distribution) $\theta = (0.2, 0.3, 0.5)$. 
\end{definition}

Our last working example is a $77$-ary source that is inspired from the true distribution of characters in password databases.
\begin{definition}[string-source $S_{77}$]
Let $S_{77}$ refer to a memoryless string-source alphabet size $|\mc{X}| =77$ where the parameter vector (categorical distribution) is inspired from the empirical distribution of characters in passwords from~\cite{distribution-passwords}. 
\end{definition}

\begin{figure*}
\centering
\includegraphics[width=0.5\textwidth]{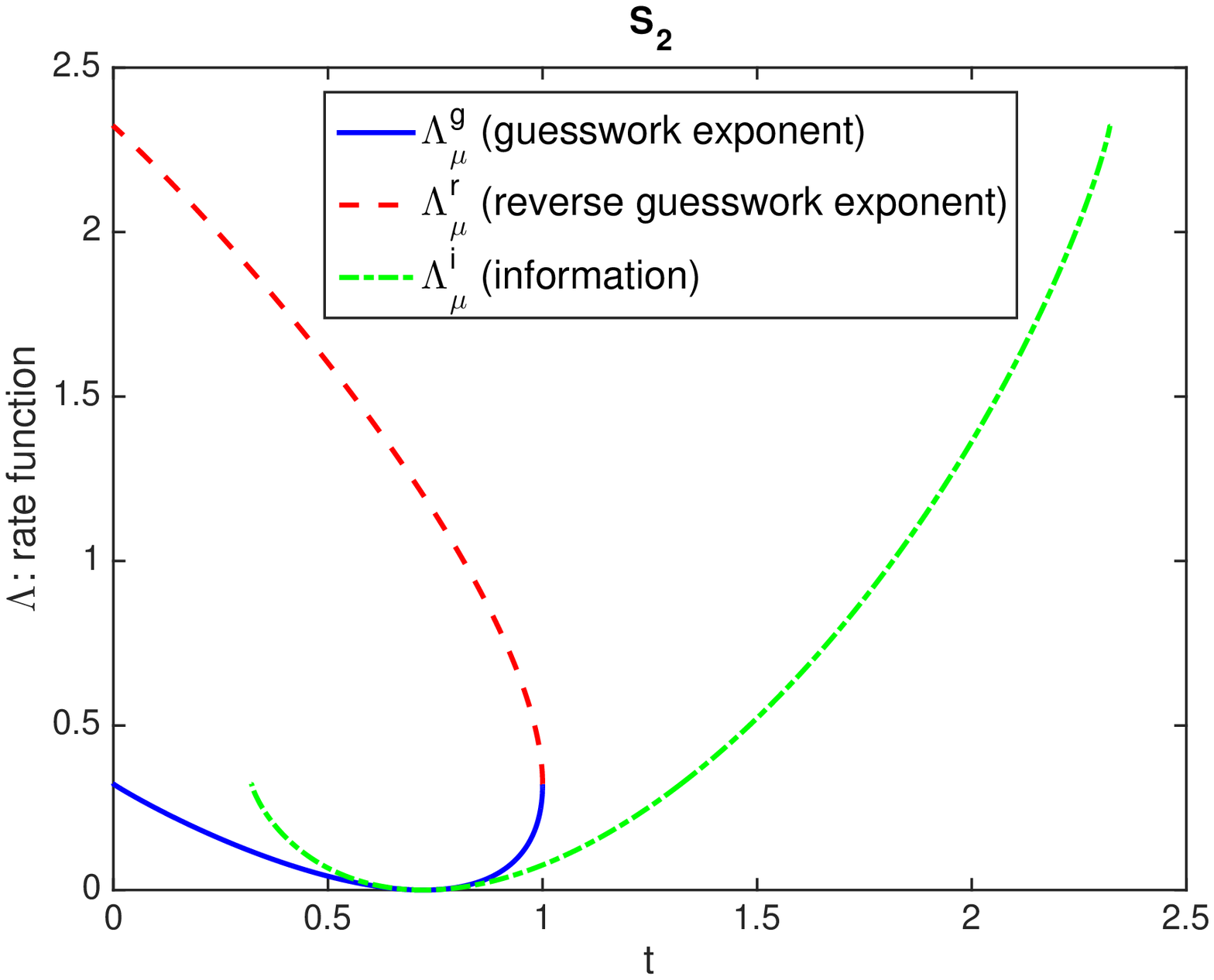}\\
\includegraphics[width=0.5\textwidth]{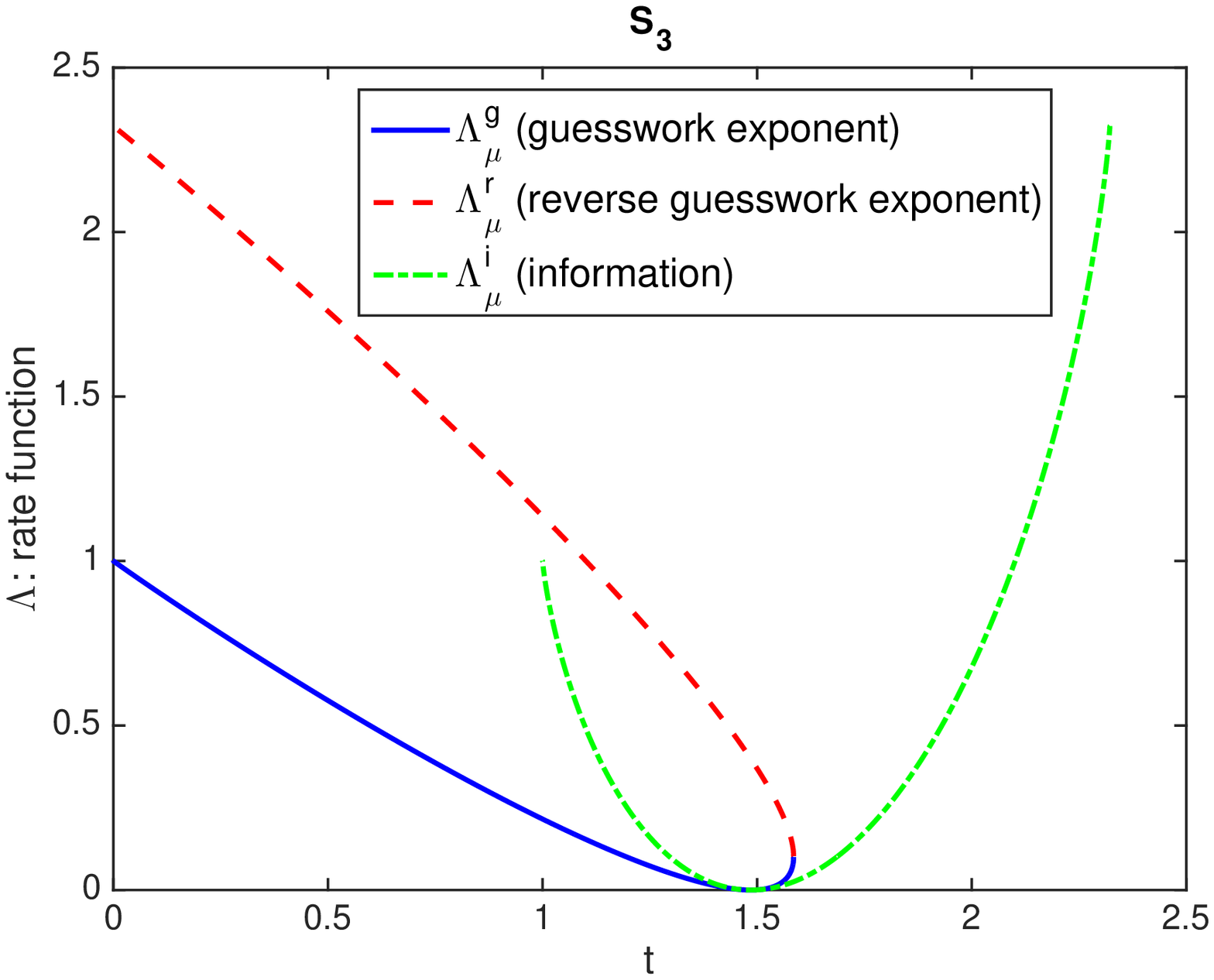}
\includegraphics[width=0.5\textwidth]{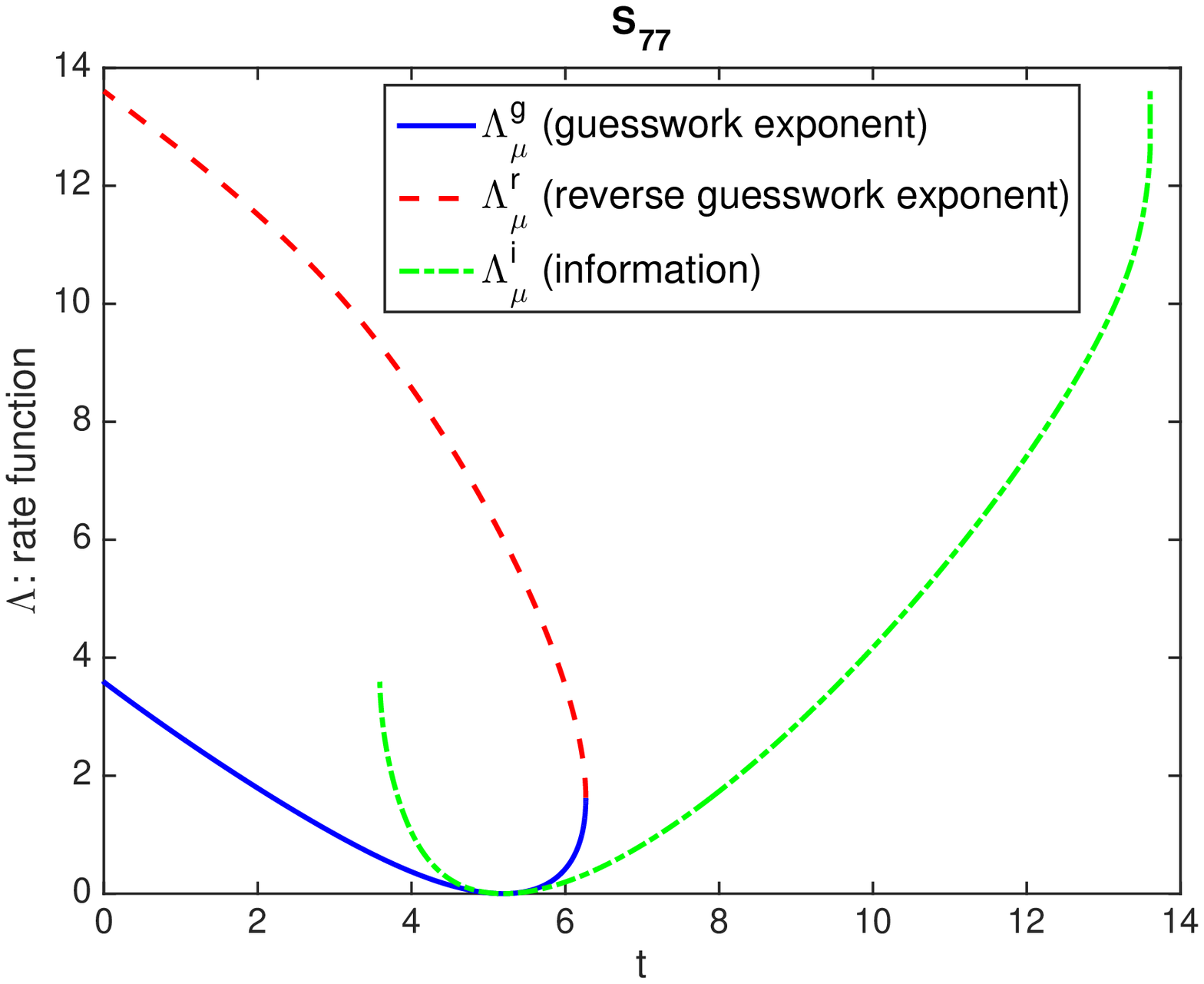}\\
\caption{The LDP rate function for $S_2$, $S_3$, and $S_{77}$.  The solid blue curve represents $J_{\mub}^g$; the dashed red curve represents $J_\mub^r$; and the dotted green curve represents $J_\mub^\imath$.}
\label{fig:LDP-rate}
\end{figure*}

We let $X^n$ be a random string of length $n$ that will be drawn from the string-source $S_i$ where $i$ will be clear from the context. The approximations of the probability are plotted in Fig.~\ref{fig:guesswork-approx1}, where we have chosen the sequence length to be $n=8$. As can be seen the approximations in~\eqref{eq:approx1} and~\eqref{eq:approx22} collectively approximate well the distribution of guesswork for different string-sources with very good accuracy even for small sequence lengths.
Note that it is straightforward to calculate the formulas provided in Approximation~\ref{approx1} for any memoryless process by sweeping $\alpha$ over the real line.

In Fig.~\ref{fig:LDP-rate}, we further plot the rate functions $J_\mub^g$, $J_\mub^r$, and $J_\mub^\imath$ for string-sources $S_2$, $S_3$, and $S_{77}$. As can be seen, $J_\mub^g(t)$ and $J_\mub^\imath(t)$ are convex functions of $t$ whereas $J_\mub^r(t)$ is a concave function of $t$ (as has already been proved). The rest of the properties of the rate functions could also be verified numerically and are observable from the plots.

\section{Concluding Remarks and Future Work}
\label{sec:conclusion}
Guesswork is an important element of information theory that is intimately related to famous problems in information theory, such as the distribution of the length of one-shot source coding, the computational complexity of sequential decoding, the probability of error in list decoding, and the quantification of computational security against brute-force attacks.
In this paper, we defined the tilt operation, leading to derivation of tight bounds and accurate approximations to the distribution, and the large deviations behavior of guesswork. Using the tilt operation, we characterized the tilted family of a string-source and established several main properties of the tilted family. In particular, we showed that the set of all string-sources that result in the same ordering of all strings from the most likely to the least likely coincides with the tilted family of the string-source. We also used the tilted family of the string-source to characterize the large deviations behavior of the logarithm of forward/reverse guesswork and information. Finally, we provided an approximation formula for the PMF of guesswork and established its effectiveness using several numerical examples.

While this paper was focused on memoryless string-sources, future work would naturally consider a richer set of stationary ergodic string-sources, such as Markov or hidden Markov processes, or other mixing processes.  This research direction appears promising, as empirically comparing actual distributions and the approximations that are obtained by applying our approximation formulas to Markov or hidden Markov string-sources yield excellent concordance.   To illustrate this point, let us define two string-sources as follows.
\begin{definition}[string-source $S^M_3$]
Let $S^M_3$ refer to a first order Markov source on alphabet $\mc{X} = \{a,b,c \}$ of size $|\mc{X}| =3$ with transition matrix $T^M_3$ given by
$$
T^M_3 = \left(
\begin{array}{lll}
0.7 & 0.1 & 0.2\\
      0.2 &  0.3 & 0.5\\
      0.3 & 0.6 & 0.1
\end{array}
\right),
$$
and where the initial distribution is chosen to be the stationary distribution of the chain.
\end{definition}

\begin{definition}[string-source $S^{H}_3$]
Let $S_3^H$ refer to a hidden Markov string-source on alphabet $\mc{X} = \{a,b,c \}$ of size $|\mc{X}| =3$ with hidden states $d$ and $e$, with $2 \times 2$ transition matrix $T^H_3$ given by
$$
T^H_3 = \left(
\begin{array}{ll}
0.8 & 0.2 \\
0.4 &  0.6
\end{array}
\right)
$$
and the emission probabilities are grouped in the matrix $E_3^H$ below.
$$
E_3^H = 
 \left(
\begin{array}{lll}
0.1 & 0.3 & 0.6\\
0.2 & 0.6 & 0.2\\
\end{array}
\right).
$$
\end{definition}

\begin{figure*}
\centering
\includegraphics[width=0.495\textwidth]{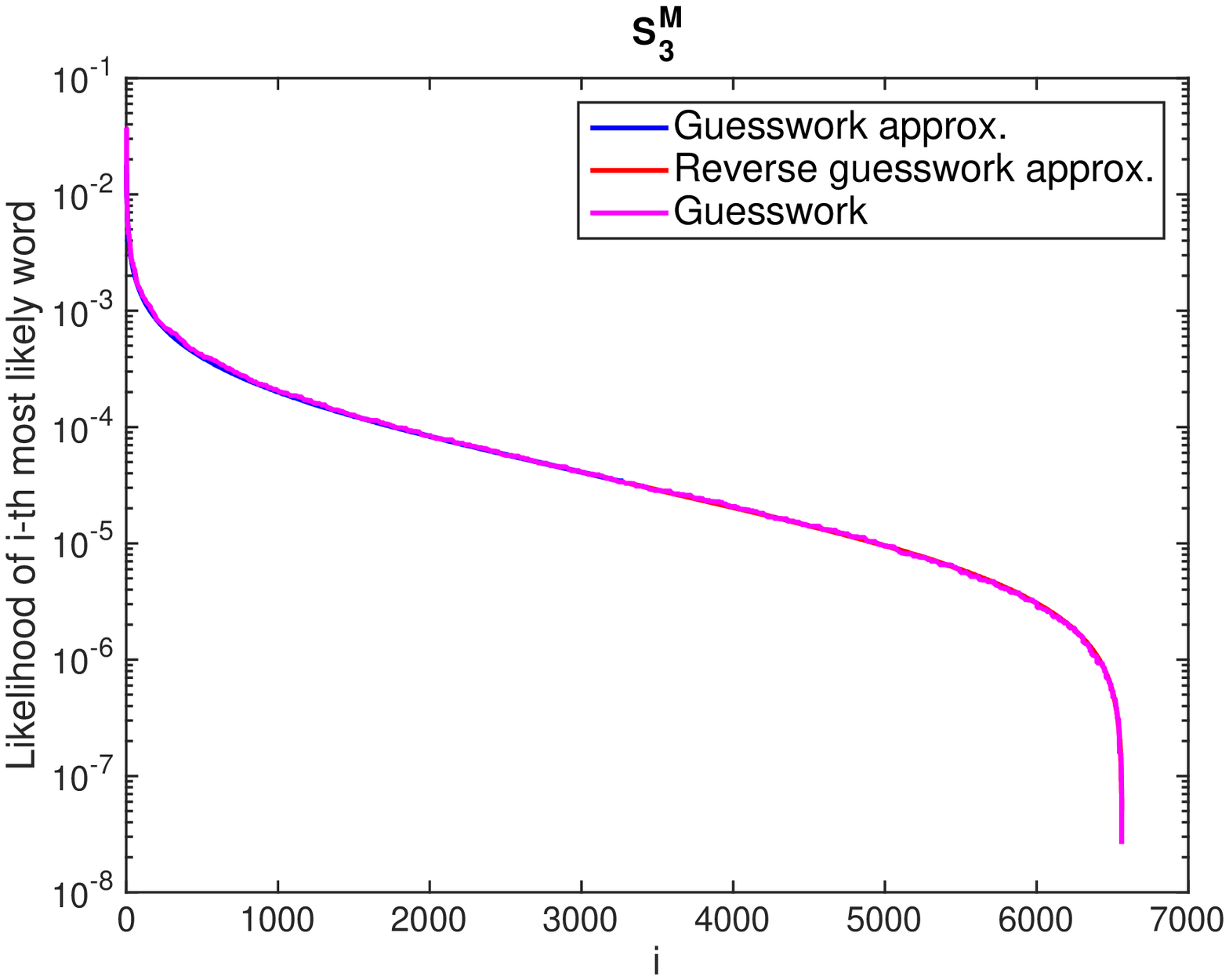}
\includegraphics[width=0.495\textwidth]{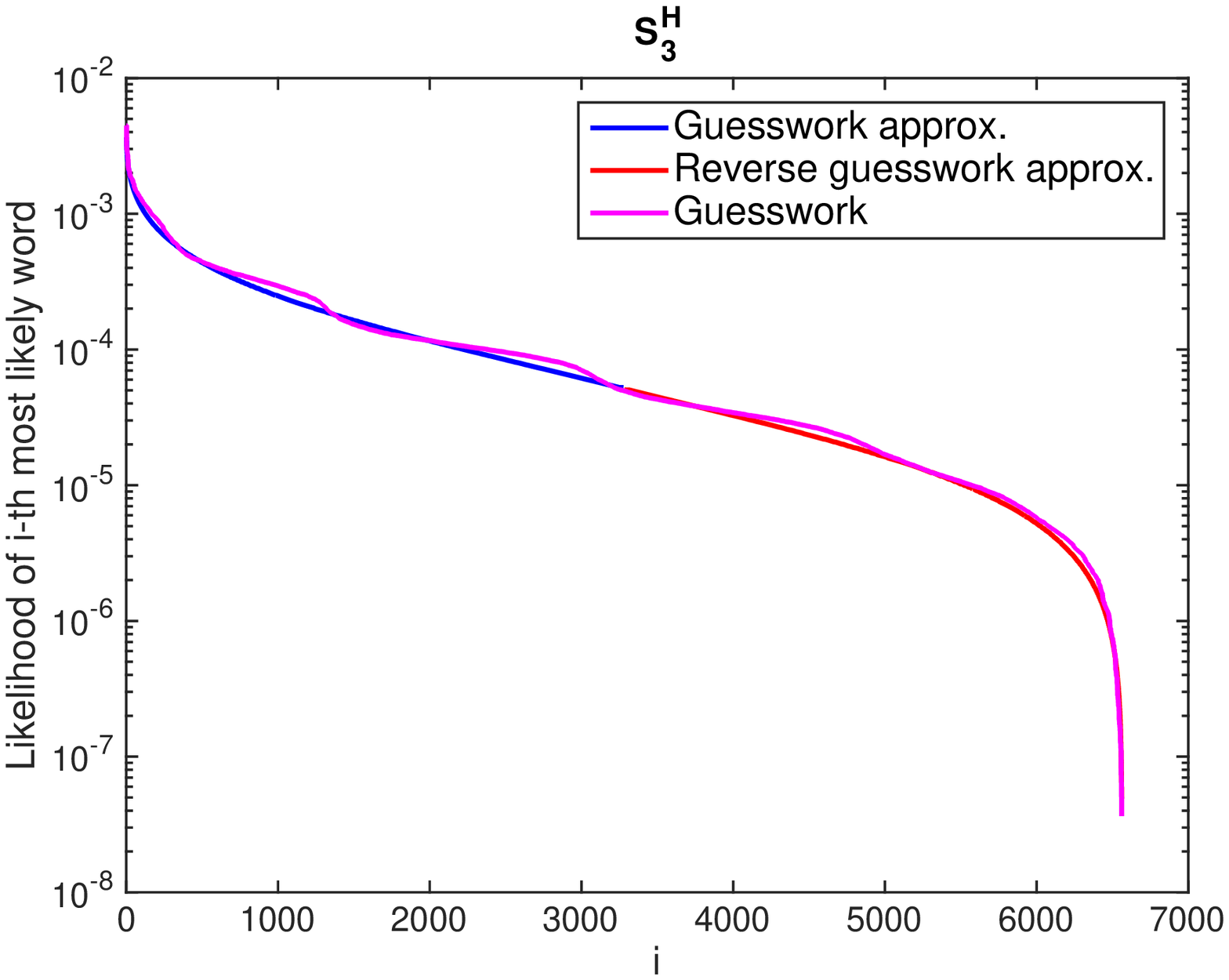}
\caption{The approximation on the distribution of the guesswork for string-sources $S_3^M$ and $S_3^H$. The string length is $n=8$. The pink staircase represents the probability mass function of guesswork on the $3^8$ strings; the blue curve is the approximation derived from the forward path in~\eqref{eq:approx1}; and the red curve is the approximation derived from the reverse path in~\eqref{eq:approx22}.}
\label{fig:guesswork-approx-conclusion}
\end{figure*}

In Fig.~\ref{fig:guesswork-approx-conclusion}, we use the formulas~\eqref{eq:approx1} and~\eqref{eq:approx22} to provide an approximation on the distribution of guesswork. Note that in this case we have used the entropy and varentropy of the words rather than the rates in order to obtain the approximation. We observe that the formulas provide accurate approximations on the distribution of guesswork even for a hidden Markov source with infinite memory, and we conclude that the generalizing techniques developed in this paper is a direction for further work that is solidly motivated by empirical evidence. Moreover, from a heuristic perspective, our guesswork approximation techniques  may provide some indicative information regarding guesswork for more sources beyond memoryless string-sources, even though the analysis currently does not provide guarantees.

\section*{Acknowledgement}
The authors are thankful to Erdal Ar{\i}kan (Bilkent University) for informative discussions about guesswork and sequential decoding, Andrew Barron (Yale University) for discussions that led to the improvement of the approximations in the small sequence length regime, Mokshay Madiman (University of Delaware) for discussions on the connections with the Shannon-McMillan-Breiman Theorem,  Ali Makhdoumi ({Duke University}) for several discussions about the properties of exponential families and careful reading of an earlier version of this paper,  and Jossy Sayir (University of Cambridge) for suggesting the symmetric plotting of the probability simplex on an equilateral triangle.
\bibliographystyle{IEEEtran}
\bibliography{phd_thesis_bib}

% Generated by IEEEtran.bst, version: 1.14 (2015/08/26)
\begin{thebibliography}{10}
\providecommand{\url}[1]{#1}
\csname url@samestyle\endcsname
\providecommand{\newblock}{\relax}
\providecommand{\bibinfo}[2]{#2}
\providecommand{\BIBentrySTDinterwordspacing}{\spaceskip=0pt\relax}
\providecommand{\BIBentryALTinterwordstretchfactor}{4}
\providecommand{\BIBentryALTinterwordspacing}{\spaceskip=\fontdimen2\font plus
\BIBentryALTinterwordstretchfactor\fontdimen3\font minus
  \fontdimen4\font\relax}
\providecommand{\BIBforeignlanguage}[2]{{%
\expandafter\ifx\csname l@#1\endcsname\relax
\typeout{** WARNING: IEEEtran.bst: No hyphenation pattern has been}%
\typeout{** loaded for the language `#1'. Using the pattern for}%
\typeout{** the default language instead.}%
\else
\language=\csname l@#1\endcsname
\fi
#2}}
\providecommand{\BIBdecl}{\relax}
\BIBdecl

\bibitem{Allerton15_guesswork}
A.~Beirami, R.~Calderbank, M.~Christiansen, K.~Duffy, A.~Makhdoumi, and
  M.~M\'edard, ``A geometric perspective on guesswork,'' in \emph{53rd Annual
  Allerton Conference (Allerton)}, Oct. 2015.

\bibitem{sequential_decoding}
I.~Jacobs and E.~Berlekamp, ``A lower bound to the distribution of computation
  for sequential decoding,'' \emph{IEEE Trans. Inf. Theory}, vol.~13, no.~2,
  pp. 167--174, April 1967.

\bibitem{massey94}
J.~L. Massey, ``Guessing and entropy,'' in \emph{1994 IEEE International
  Symposium on Information Theory Proceedings}, 1994, p. 204.

\bibitem{Arikan-guesswork}
E.~Ar{\i}kan, ``An inequality on guessing and its application to sequential
  decoding,'' \emph{IEEE Trans. Inf. Theory}, vol.~42, no.~1, pp. 99--105, Jan.
  1996.

\bibitem{sullivan-markov}
D.~Malone and W.~G. Sullivan, ``Guesswork and entropy,'' \emph{IEEE Trans. Inf.
  Theory}, vol.~50, no.~3, pp. 525--526, Mar. 2004.

\bibitem{sullivan-stationary}
C.~E. Pfister and W.~G. Sullivan, ``Renyi entropy, guesswork moments, and large
  deviations,'' \emph{IEEE Trans. Inf. Theory}, vol.~50, no.~11, pp.
  2794--2800, Nov. 2004.

\bibitem{sason-guesswork}
I.~Sason and S.~Verd{\'u}, ``Improved bounds on lossless source coding and
  guessing moments via r{\'e}nyi measures,'' \emph{IEEE Transactions on
  Information Theory}, vol.~64, no.~6, pp. 4323--4346, 2018.

\bibitem{guesswork-distortion}
E.~Ar{\i}kan and N.~Merhav, ``Guessing subject to distortion,'' \emph{IEEE
  Trans. Inf. Theory}, vol.~44, no.~3, pp. 1041--1056, May 1998.

\bibitem{ISIT15_guesswork}
A.~Beirami, R.~Calderbank, K.~Duffy, and M.~M\'edard, ``Quantifying
  computational security subject to source constraints, guesswork and
  inscrutability,'' in \emph{2015 IEEE International Symposium on Information
  Theory Proceedings (ISIT)}, Jun. 2015.

\bibitem{guessing-revisited}
M.~K. Hanawal and R.~Sundaresan, ``Guessing revisited: A large deviations
  approach,'' \emph{IEEE Trans. Inf. Theory}, vol.~57, no.~1, pp. 70--78, Jan.
  2011.

\bibitem{duffy-LDP}
M.~M. Christiansen and K.~R. Duffy, ``Guesswork, large deviations, and
  {S}hannon entropy,'' \emph{IEEE Trans. Inf. Theory}, vol.~59, no.~2, pp.
  796--802, Feb. 2013.

\bibitem{Duffy-CCT}
K.~R. Duffy, J.~Li, and M.~M{\'e}dard, ``Guessing noise, not code-words,'' in
  \emph{2018 IEEE International Symposium on Information Theory (ISIT)}.\hskip
  1em plus 0.5em minus 0.4em\relax IEEE, 2018, pp. 671--675.

\bibitem{Elias-list-1}
P.~Elias, \emph{List decoding for noisy channels}, 1957.

\bibitem{Elias-list-2}
------, ``Error-correcting codes for list decoding,'' \emph{IEEE Trans. Inf.
  Theory}, vol.~37, no.~1, pp. 5--12, 1991.

\bibitem{Sudan-list}
M.~Sudan, ``List decoding: Algorithms and applications,'' in \emph{Theoretical
  Computer Science: Exploring New Frontiers of Theoretical Informatics}.\hskip
  1em plus 0.5em minus 0.4em\relax Springer, 2000, pp. 25--41.

\bibitem{Gallager-book}
R.~G. Gallager, \emph{Information theory and reliable communication}.\hskip 1em
  plus 0.5em minus 0.4em\relax Springer, 1968, vol.~2.

\bibitem{Merhav-list}
N.~Merhav, ``List decoding---{R}andom coding exponents and expurgated
  exponents,'' \emph{IEEE Trans. Inf. Theory}, vol.~60, no.~11, pp. 6749--6759,
  2014.

\bibitem{multi-user}
M.~Christiansen, K.~Duffy, F.~du~Pin~Calmon, and M.~M\'edard, ``Multi-user
  guesswork and brute force security,'' \emph{IEEE Trans. Inf. Theory},
  vol.~61, no.~12, pp. 6876--6886, Dec 2015.

\bibitem{ISIT13_Mark}
M.~M. Christiansen, K.~R. Duffy, F.~du~Pin~Calmon, and M.~M{\'e}dard, ``Brute
  force searching, the typical set and guesswork,'' in \emph{2013 IEEE
  International Symposium on Information Theory Proceedings (ISIT)}, July 2013,
  pp. 1257--1261.

\bibitem{ASILOMAR13}
------, ``Guessing a password over a wireless channel (on the effect of noise
  non-uniformity),'' in \emph{2013 Asilomar Conference on Signals, Systems and
  Computers}, 2013, pp. 51--55.

\bibitem{sundaresan-universal}
R.~Sundaresan, ``Guessing under source uncertainty,'' \emph{IEEE Trans. Inf.
  Theory}, vol.~53, no.~1, pp. 269--287, Jan. 2007.

\bibitem{IT_Kosut_one2one}
O.~Kosut and L.~Sankar, ``Asymptotics and non-asymptotics for universal
  fixed-to-variable source coding,'' \emph{arXiv preprint arXiv:1412.4444},
  2014.

\bibitem{Wyner72}
A.~D. Wyner, ``An upper bound on the entropy series,'' \emph{Information and
  Control}, vol.~20, no.~2, pp. 176--181, 1972.

\bibitem{Alon_Orlitsky_one2one}
N.~Alon and A.~Orlitsky, ``A lower bound on the expected length of one-to-one
  codes,'' \emph{IEEE Trans. Inf. Theory}, vol.~40, no.~5, pp. 1670--1672,
  Sept. 1994.

\bibitem{courtade-verdu}
T.~Courtade and S.~Verd{\'u}, ``Cumulant generating function of codeword
  lengths in optimal lossless compression,'' in \emph{2014 IEEE International
  Symposium on Information Theory (ISIT)}, July 2014, pp. 2494--2498.

\bibitem{Szpankowski_one2one}
W.~Szpankowski, ``A one-to-one code and its anti-redundancy,'' \emph{IEEE
  Trans. Inf. Theory}, vol.~54, no.~10, pp. 4762--4766, Oct. 2008.

\bibitem{Szpankowski2011}
W.~Szpankowski and S.~Verd\'u, ``Minimum expected length of fixed-to-variable
  lossless compression without prefix constraints,'' \emph{IEEE Trans. Inf.
  Theory}, vol.~57, no.~7, pp. 4017--4025, Jul. 2011.

\bibitem{Kontoyiannis_one2one}
I.~Kontoyiannis and S.~Verd\'u, ``Optimal lossless data compression:
  Non-asymptotics and asymptotics,'' \emph{IEEE Trans. Inf. Theory}, vol.~60,
  no.~2, pp. 777--795, Feb. 2014.

\bibitem{ITW14}
A.~Beirami and F.~Fekri, ``Fundamental limits of universal lossless one-to-one
  compression of parametric sources,'' in \emph{2014 IEEE Information Theory
  Workshop (ITW '14)}, Nov. 2014, pp. 212--216.

\bibitem{strassen}
V.~Strassen, ``Asymptotische absch{\"a}tzungen in shannons informations
  theorie,'' in \emph{Trans. Third Prague Conf. Inf. Theory}, 1962, pp.
  689--723.

\bibitem{Renyi-entropy}
A.~R\'enyi, ``On measures of entropy and information,'' in \emph{Fourth
  Berkeley Symposium in Mathematical Statistics}, 1961.

\bibitem{information-geometry}
I.~Csisz{\'a}r and P.~C. Shields, \emph{Information theory and statistics: A
  tutorial}.\hskip 1em plus 0.5em minus 0.4em\relax Now Publishers Inc, 2004.

\bibitem{lizhong-allerton-2006}
S.~Borade and L.~Zheng, ``I-projection and the geometry of error exponents,''
  in \emph{Proceedings of the Forty-Fourth Annual Allerton Conference on
  Communication, Control, and Computing, Sept 27-29}, 2006.

\bibitem{meyn-entropy}
J.~Huang, C.~Pandit, S.~Meyn, M.~M\'edard, and V.~Veeravalli, ``Entropy,
  inference, and channel coding,'' in \emph{Wireless Communications}.\hskip 1em
  plus 0.5em minus 0.4em\relax Springer, 2007, pp. 99--124.

\bibitem{Barron_Cover_91}
A.~R. Barron and T.~M. Cover, ``Minimum complexity density estimation,''
  \emph{IEEE Trans. Inf. Theory}, vol.~37, no.~4, pp. 1034--1054, Jul. 1991.

\bibitem{Clarke_Barron}
B.~Clarke and A.~Barron, ``Information-theoretic asymptotics of {B}ayes
  methods,'' \emph{IEEE Trans. Inf. Theory}, vol.~36, no.~3, pp. 453 --471, May
  1990.

\bibitem{dembo-book}
A.~Dembo and O.~Zeitouni, \emph{Large deviations techniques and
  applications}.\hskip 1em plus 0.5em minus 0.4em\relax Springer Science \&
  Business Media, 2009, vol.~38.

\bibitem{cover-book}
T.~M. Cover and J.~A. Thomas, \emph{Elements of information theory}.\hskip 1em
  plus 0.5em minus 0.4em\relax John Wiley and Sons, 2006.

\bibitem{mokshay-AEP2}
S.~Bobkov and M.~Madiman, ``Concentration of the information in data with
  log-concave distributions,'' \emph{The Annals of Probability}, vol.~39,
  no.~4, pp. 1528--1543, 2011.

\bibitem{mokshay-AEP}
S.~Bobkov and M.~M. Madiman, ``An equipartition property for high-dimensional
  log-concave distributions.'' in \emph{Allerton Conference}, 2012, pp.
  482--488.

\bibitem{Shannon-McMillan}
D.~Ornstein and B.~Weiss, ``Entropy and data compression schemes,'' \emph{IEEE
  Trans. Inf. Theory}, vol.~39, no.~1, pp. 78--83, Jan. 1993.

\bibitem{Barron-AEP}
A.~R. Barron, ``{The strong ergodic theorem for densities: generalized
  Shannon-McMillan-Breiman theorem},'' \emph{The Annals of Probability}, pp.
  1292--1303, 1985.

\bibitem{frank-note}
F.~R. Kschischang, ``The complementary error function,'' [Online, April
  2017]~\url{https://www.comm.utoronto.ca/frank/notes/erfc.pdf}.

\bibitem{distribution-passwords}
``Character occurrence in passwords,'' [Online, June
  2011]~\url{https://csgillespie.wordpress.com/2011/06/16/character-occurrence-in-passwords}.

\end{thebibliography}

\begin{IEEEbiographynophoto}{Ahmad Bierami} is a research scientist at Electronic Arts Digital Platform -- Data \& AI,  leading fundamental research and development on training AI agents in multi-agent systems. His research interests broadly include AI, machine learning, statistics, information theory, and networks. Prior to joining EA in 2018, he held postdoctoral fellow positions at Duke, MIT, and Harvard. He received the BS degree in 2007 from Sharif University of Technology, Iran, and the PhD degree in 2014 from the Georgia Institute of Technology, in electrical and computer engineering. He is the recipient of the 2015 Sigma Xi Best PhD Thesis Award from Georgia Tech.
\end{IEEEbiographynophoto}

\begin{IEEEbiographynophoto}{Robert Calderbank} (M'89--SM'97--F'98) received the BSc degree in 1975 from Warwick University,
England, the MSc degree in 1976 from Oxford University, England, and the PhD degree in 1980 from the
California Institute of Technology, all in mathematics.

Dr. Calderbank is Professor of Electrical and Computer Engineering at Duke University where he directs
the Information Initiative at Duke (iiD). Prior to joining Duke in 2010, Dr. Calderbank was Professor of
Electrical Engineering and Mathematics at Princeton University where he directed the Program in
Applied and Computational Mathematics. Prior to joining Princeton in 2004, he was Vice President for
Research at AT\&T, responsible for directing the first industrial research lab in the world where the
primary focus is data at scale. At the start of his career at Bell Labs, innovations by Dr. Calderbank were
incorporated in a progression of voiceband modem standards that moved communications practice
close to the Shannon limit. Together with Peter Shor and colleagues at AT\&T Labs he developed the
mathematical framework for quantum error correction. He is a co-inventor of space-time codes for
wireless communication, where correlation of signals across different transmit antennas is the key to
reliable transmission.

Dr. Calderbank served as Editor in Chief of the IEEE TRANSACTIONS ON INFORMATION THEORY from
1995 to 1998, and as Associate Editor for Coding Techniques from 1986 to 1989. He was a member of
the Board of Governors of the IEEE Information Theory Society from 1991 to 1996 and from 2006 to
2008. Dr. Calderbank was honored by the IEEE Information Theory Prize Paper Award in 1995 for his
work on the Z4 linearity of Kerdock and Preparata Codes (joint with A.R. Hammons Jr., P.V. Kumar, N.J.A.
Sloane, and P. Sole), and again in 1999 for the invention of space-time codes (joint with V. Tarokh and N.
Seshadri). He has received the 2006 IEEE Donald G. Fink Prize Paper Award, the IEEE Millennium Medal,
the 2013 IEEE Richard W. Hamming Medal, and the 2015 Shannon Award. He was elected to the US
National Academy of Engineering in 2005.
\end{IEEEbiographynophoto}

\begin{IEEEbiographynophoto}{Mark M. Christiansen} is an assistant manager in the quantitative risk team in Grant Thornton. He received his B.A. (mod) in mathematics from Trinity College Dublin and his PhD in mathematics at Hamilton Institute in Maynooth. His interests include Probability Theory, Information Theory and their use in finance and security.
\end{IEEEbiographynophoto}

\begin{IEEEbiographynophoto}{Ken R. Duffy} is a Professor at Maynooth University where he is currently the Director of the Hamilton Institute. He received the B.A.(mod) and Ph.D. degrees in mathematics from the Trinity College Dublin. His primary research interests are in probability and statistics, and their applications in science and engineering.
\end{IEEEbiographynophoto}

\begin{IEEEbiographynophoto}{Muriel M\'edard} is the Cecil H. Green Professor in the Electrical Engineering and Computer Science (EECS) Department at MIT and leads the Network Coding and Reliable Communications Group at the Research Laboratory for Electronics at MIT. She has co-founded three companies to commercialize network coding, CodeOn, Steinwurf and Chocolate Cloud. She has served as editor for many publications of the Institute of Electrical and Electronics Engineers (IEEE), of which she was elected Fellow, and she has served as Editor in Chief of the IEEE Journal on Selected Areas in Communications. She was President of the IEEE Information Theory Society in 2012, and served on its board of governors for eleven years. She has served as technical program committee co-chair of many of the major conferences in information theory, communications and networking. She received the 2009 IEEE Communication Society and Information Theory Society Joint Paper Award, the 2009 William R. Bennett Prize in the Field of Communications Networking, the 2002 IEEE Leon K. Kirchmayer Prize Paper Award, the 2018 ACM SIGCOMM Test of Time Paper Award and several conference paper awards. She was co-winner of the MIT 2004 Harold E. Edgerton Faculty Achievement Award, received the 2013 EECS Graduate Student Association Mentor Award and served as Housemaster for seven years. In 2007 she was named a Gilbreth Lecturer by the U.S. National Academy of Engineering. She received the 2016 IEEE Vehicular Technology James Evans Avant Garde Award, the 2017 Aaron Wyner Distinguished Service Award from the IEEE Information Theory Society and the 2017 IEEE Communications Society Edwin Howard Armstrong Achievement Award.
\end{IEEEbiographynophoto}

\end{document}